\documentclass[hyper]{JHEP} % 10pt is ignored!
\usepackage{graphicx,amssymb,bm,latexsym,amsmath}
\usepackage{epstopdf}
\usepackage{caption}
\usepackage{subcaption}

\newcommand\fverb{\setbox\pippobox=\hbox\bgroup\verb}
\newcommand\fverbdo{\egroup\medskip\noindent%
            \fbox{\unhbox\pippobox}\ }

\newcommand\fverbit{\egroup\item[\fbox{\unhbox\pippobox}]}
\newbox\pippobox
\title{Spiky strings in $\varkappa$-deformed $AdS$}
\author{Aritra Banerjee\\
Department of Physics, Indian Institute of Technology Kharagpur,\\
Kharagpur-721 302, INDIA \\
Email: \email{aritra@phy.iitkgp.ernet.in}}
\author{Soumya Bhattacharya\\
Centre for Theoretical Studies, Indian Institute of Technology Kharagpur,\\
Kharagpur-721 302, INDIA \\
Email: \email{soumya557@gmail.com}}
\author{Kamal L. Panigrahi\\
Department of Physics, Indian Institute of Technology Kharagpur,\\
Kharagpur-721 302, INDIA   \\

Email: \email{panigrahi@phy.iitkgp.ernet.in}}
\abstract{We study rigidly rotating strings in
$\varkappa$-deformed $AdS$ background. We probe this classically
integrable background with `spiky' strings and analyze the string
profiles in the large charge limit systematically. We also discuss
the dispersion relation among the conserved charges for these
solutions in long string limit.}

\begin{document}
\section{Introduction}
Integrability has been proved to be one of the most useful tools
in studying string spectrum in various semisymmetric superspaces
\cite{Zarembo:2010sg}\footnote{For an introduction to integrability 
in AdS/CFT one can see \cite{Beisert:2010jr}.}. The integrable structure
on both sides of
the celebrated AdS/CFT correspondence \cite{Maldacena:1997re} has
opened up new areas in the study of string theory, among others,
for example the type IIB superstring theory on $AdS_5\times S^5$
has been shown to be described as supercoset sigma model
\cite{Metsaev:1998it} and integrability of AdS has been explored
in \cite{Bena:2003wd}. It has been shown that in the large angular
momentum or large R-charge limit, both sides of the AdS/CFT
duality are easier to probe \cite{Pohlmeyer:1975nb,Minahan:2002rc,
Tseytlin:2004xa,Hayashi:2007bq,Okamura:2008jm}. It was further
noticed that the spectrum of semiclassical spinning strings in
this limit can be related to the anomalous dimension of the ${\cal
N} = 4$ Super Yang-Mills (SYM), and this can be derived from the
relation between conserved charges of the rigidly rotating strings
in $AdS_5 \times S^5$. This comes from the amazing proposal of
\cite{Minahan:2002ve,Beisert:2003yb} which relates the dilatation
operator in  ${\cal N} = 4$ Super Yang-Mills to the Hamiltonian of
an integrable Heisenberg $SU(2)$ type spin-chain system. Inspired
by this, many studies of rotating strings on $AdS$ and
asymptotically $AdS$ spaces have come up, for example in
\cite{misc1,misc2,misc5,misc6,misc7,misc8,misc9}.

The integrable system description of this duality opens up
possibility of illuminating the gauge theories dual to integrable
deformations of the $AdS_5\times S^5$ case. For many cases, these
kind of deformations have been achieved by T-dualizing the
existing low-dimensional bosonic sigma models
\cite{Lunin:2005jy,Frolov:2005ty,Frolov:2005dj,
Alday05,Ricci:2007eq,Beisert:2008iq}. Classical strings on these
backgrounds are also studied in, for example,
\cite{deform1,deform2}. Contrary to these models, a novel
one-parameter deformation of the bosnic sigma model was
constructed in \cite{Delduc:2013qra} following a few earlier
proposals on Yang-Baxter sigma models put forward by Klimcik 
\cite{Klimcik:2002zj,Klimcik:2008eq,Klimcik:2014bta}
\footnote{Yang-Baxter sigma models based on classical Yang-Baxter
equation (CYBE) were recently studied at length in, e.g. \cite{Kawaguchi:2014qwa}.
In \cite{Matsumoto:2014nra} it was further shown that the TsT type deformations mentioned
earlier also  belong to the class of Yang-Baxter type deformations.}.
This real deformation parameter is often called
$\varkappa$ or $\eta$ with $\varkappa = \frac{2\eta}{1-\eta^2}$ ,
where $\varkappa\in[0,\infty)$. The foundations of the integrable
structure of string theory as described by this deformed sigma
model have been presented in
\cite{ABF,HRT,Arutynov:2014ota,Arutyunov:2014cra,Delduc:2014kha,
Hollowood:2014rla,Arutyunov:2014jfa}.

The new deformed model has the
bosonic symmetry group reduced from $SO(2,4) \times SO(6)$  for
$AdS_5\times S^5$ to its  Cartan subgroup $[U(1)]^3 \times
[U(1)]^3$, which makes  most of the symmetry of  the original
space hidden or sometimes called ``$q$-deformed". Indeed the
deformed integrable sigma model has a $q$-deformed $PSU(2,2\lvert 4)$
as the symmetry group \cite{Delduc:2014kha}, with $q$ as a function of
$\eta$. This model seemingly has no
manifest space-time supersymmetry but the $q$-deformed symmetry
group suggests that the space-time supersymmetry can be represented in
a non-trivial way. The exact 10d
metric and the associated Neveu-Schwarz B-field were found in
\cite{ABF} and various consistent truncations were also
discussed in \cite{Hoare:2014pna}. The corresponding deformed full type IIB supergravity
solutions for the subset $AdS_2 \times S^2$ and $AdS_3 \times S^3$
have been found out recently in \cite{Lunin:2014tsa}. It has been
shown that the solutions depend on a free parameter which in turn
is a function of the deformation parameter $\varkappa$. In a
related development, the two parameter and three parameter
generalizations of this deformed supercoset sigma model have been
outlined in \cite{Hoare:2014oua}. It seems that there is a
singularity surface associated with these deformed background
metrics, which was addressed in detail in \cite{Kameyama:2014vma}.
Various minimal surfaces and cusped Wilson loops in this 
background were studied in \cite{Kameyama:2014via, Bai:2014pya}.
Also more recently, some three point correlation functions have
been studied in \cite{Bozhilov:2015kya}. The gauge theory dual
of this solution, if any, needs to be found fully.

Now, given the integrable nature of this deformation, it is
natural to look for rigidly rotating string solutions and explore
whether there can be a spin-chain like underlying system or not.
In this connection, in a subspace of the deformed $AdS_5\times
S^5$, the so called giant magnon and single spike solutions of the
string have been studied along with the finite size corrections
\cite{Arutynov:2014ota, Kameyama:2014vma, Khouchen:2014kaa, Ahn:2014aqa,
Banerjee:2014bca}. Despite the complex  structure of
the giant magnon solution on a deformed $\mathbb{R} \times S^2$
subspace, it clearly had retained the periodicity in magnon momenta $p$.
It was also shown in the limit $\varkappa \rightarrow 0$ , the relation
reduces to the form of usual giant magnon dispersion relation
proposed by Hofman-Maldacena (HM) \cite{Hofman:2006xt}. Also it
was shown in \cite{Banerjee:2014bca} that the usual single spike
strings \cite{Ishizeki:2007we} and the giant magnon solutions on
the deformed sphere can be derived as two limits of a single
setup. The deformed Neumann and the Neumann-Rosochatius
systems for spinning-strings have been described in
\cite{Arutyunov:2014cda, Kameyama:2014vma}.
On the deformed $AdS$, the usual folded GKP like string
solutions \cite{Gubser:2002tv} were discussed in
\cite{Kameyama:2014vma} and it was shown that in the `long' string
limit, the expression for cusp anomalous dimension does not reduce
to the undeformed $AdS$ case even if one takes the limit
$\varkappa \rightarrow 0$. Although in the investigation of
`pulsating' strings on $(AdS)_\varkappa$ \cite{Panigrahi:2014sia}
no such inconsistencies were found.

In the present paper, we focus on the `spiky' string profiles having multiple cusps
in the $\varkappa$ deformed $AdS$ by probing a classical rotating
string. The symmetric spiky strings on $AdS_3$ were first studied
in \cite{Kruczenski:2004wg} from the worldsheet viewpoint.
The usual dispersion relations for `long' $AdS$ spiky strings were
 constructed in \cite{Kruczenski:2004wg,Beccaria:2008tg}. A spike
(or cusp) is defined as a discontinuity in the spacelike unit
tangent vector to the string, which can also be present on a
smooth worldsheet. They correspond to single trace operators in
the dual gauge theories. Also more general spiky string solutions
in $AdS$ were described in \cite{Ryang:2006yq} in terms of a
Polyakov string with a generalized embedding ansatz for the open
string. \footnote{It is worth mentioning here that for a nice exposure to 
integrability in $AdS_3/CFT_2$ correspondence one might go 
through \cite{Sfondrini:2014via} and references therein.}

The rest of the paper is organized as follows. In section-1, we
describe rigidly rotating open string profiles with a general
embedding ansatz. After revisiting the string solutions in
\cite{Ryang:2006yq} completely and systematically classifying the profiles in
various regions of the parameter space, we discuss the same in the
$\varkappa$ deformed $AdS_3 \times S^1$. We discuss the effect of
turning on the deformation parameter on the string profiles. We
also write an approximate dispersion relation for the 2-spin spiky
strings. In section-2, we discuss the particular kind of closed
spiky strings studied in \cite{Kruczenski:2004wg} in this deformed
background and plot the symmetric string profiles for various
values of the parameters involved. We also find out the behaviour
of the conserved charges for each spike in the large spin limit.
This reduces to the same expression for $E - S$ described in
\cite{frolov} in the GKP limit, at least in the leading order approximation.
Finally in section-3, we present our conclusions and outlook.

\section{Polyakov strings and Spikes in $(AdS)_{\varkappa}$}
In this subsection we mainly discuss about rigidly rotating open
strings in the lines of \cite{Ryang:2006yq}. After revisiting the
original string solutions in $AdS_3 \times S^1$, we will
generalize it to one parameter deformed geometry.
\subsection{Classifying open strings in $AdS_3 \times S^1$}
\begin{figure}
\begin{center}
 \includegraphics[width=0.8\linewidth]{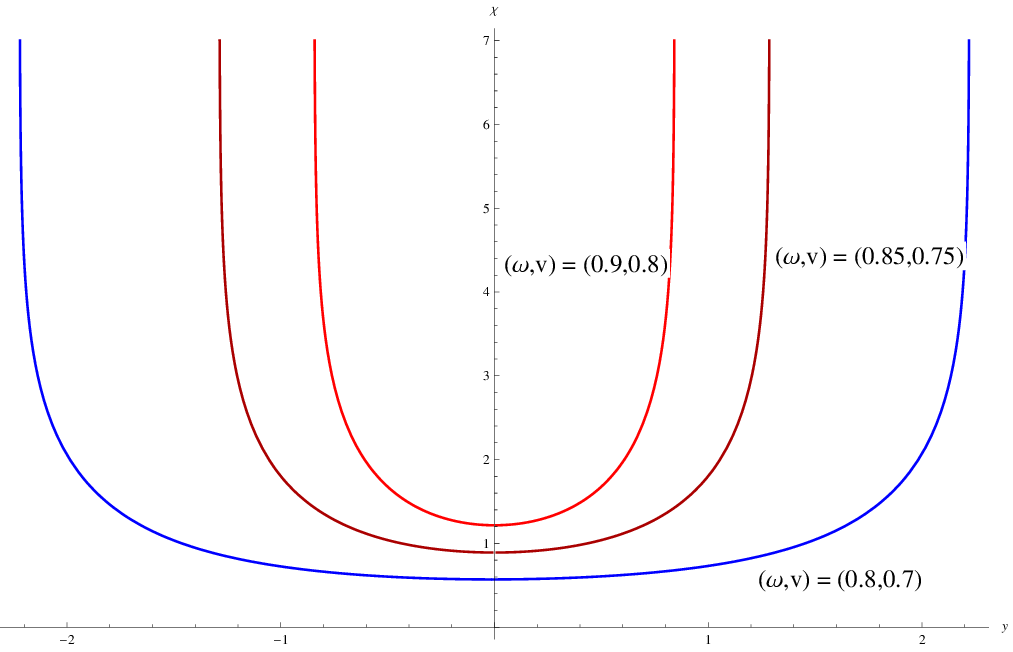}
\caption{Hanging string profiles for various values of $(\omega, v)$.}\label{fig:hangingno}
\end{center}
\end{figure}

\begin{figure}
\begin{center}
 \includegraphics[width=0.8\linewidth]{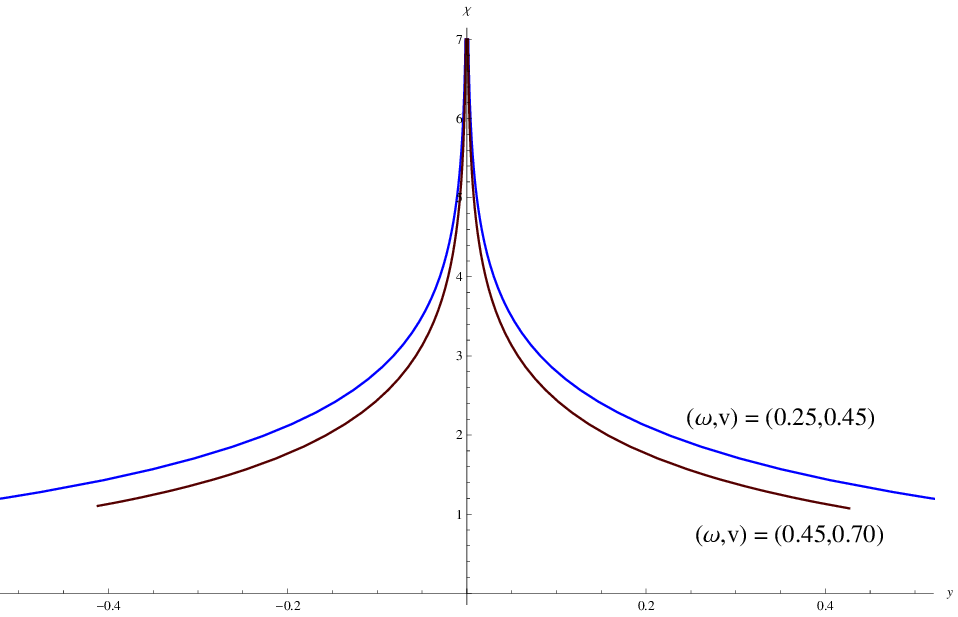}
\caption{Spiky string profiles for various values of $(\omega, v)$.}\label{fig:spikyno}
\end{center}
\end{figure}

In this section we review the main features of the multispin giant
magnon-like solutions in undeformed $AdS$ as discussed in
\cite{Ryang:2006yq} and present visual description of the
solutions. For this we start with the usual $AdS_3 \times S^3$
background given by the following metric
\begin{equation}\label{globalmetric}
ds^2 = -\cosh^2{\chi}dt^2 + d\chi^2 + \sinh^2{\chi}d\phi^2
+ \sin^2\theta d \varphi^2 + d\theta^2 + \cos^2\theta d \psi^2
\end{equation}
Since $\theta = \frac{\pi}{2}$ is a viable solution for the classical
string equations of motion for this background, we can write the
usual $AdS_3 \times S^1$ metric as
\begin{equation}\label{metric2}
ds^2 = -\cosh^2{\chi}dt^2 + d\chi^2 + \sinh^2{\chi}d\phi^2 + d \varphi^2.
\end{equation}
To describe the 2-spin giant magnon/spiky string solution we take the
rotating string embedding ansatz as follows
\begin{equation} \label{ansatz}
t=\tau+g_1(y), \quad \chi = \chi(y),\quad \phi = \omega (\tau +
g_2(y)),\quad \varphi= \mu\, \tau,
\end{equation}
where $y=\sigma-v \tau$ and $0<v<1$. The polyakov action of the
F-string is given as
\begin{equation}
S = -\frac{T}{2}\int d\sigma d\tau
[\sqrt{-\gamma}\gamma^{\alpha \beta}g_{MN}\partial_{\alpha} X^M
\partial_{\beta}X^N ]\ .   \label{action}
\end{equation}
Variation of the action (\ref{action}) with
respect to $X^M$ gives us the following equation of motion
\begin{eqnarray}
2\partial_{\alpha}(\eta^{\alpha \beta} \partial_{\beta}X^Ng_{KN})
&-& \eta^{\alpha \beta} \partial_{\alpha} X^M \partial_{\beta}
X^N\partial_K  g_{MN} =0 \ ,
\end{eqnarray}
and variation with respect to the metric gives the two Virasoro
constraints,
\begin{eqnarray}
g_{MN}(\partial_{\tau}X^M \partial_{\tau}X^N +
\partial_{\sigma}X^M \partial_{\sigma}X^N)&=&0 \ , \nonumber \\
g_{MN}(\partial_{\tau}X^M \partial_{\sigma}X^N)&=&0 \ .
\end{eqnarray}
Now the equations of motion for $t$ and $\phi$ gives
\begin{equation}
 \partial_y g_1 = \frac{1}{1 - v^2}\left[\frac{C_1}{\cosh^2\chi} - v\right],
 \\\\\\ \ \partial_y g_2 = \frac{1}{1 - v^2}\left[\frac{C_1}{\sinh^2\chi} -
 v\right] \ .
\end{equation}
 Also from the $\chi$ equation, supplemented by the boundary
 condition $\chi\prime(y)\rightarrow 0$
 as $\chi \rightarrow 0$, we get the condition $C_2 = 0$. The equation
 of motion then can be written as
 \begin{equation}
 \chi\prime(y)= \pm \frac{\sqrt{1-\omega^2}}{(1-v^2)}\tanh\chi \sqrt{\cosh^2\chi
 - \frac{C_1^2}{1-\omega^2}} \ . \label{chieq}
 \end{equation}
 Again subtracting the two Virasoro constraints we get a relation
between the various constants of the embedding ansatz
\begin{equation}
 \mu^2 - \frac{ C_1}{v} = 0 \ . \label{relation}
\end{equation}
Also by equating the $\chi$ equation with the first Virasoro
constraint we get the equation for $C_1$ as
\begin{equation}
 C_1^2 - C_1\frac{1+v^2}{v} +1 = 0,
\end{equation}
the solutions of which gives the values of $C_1$ consistent with
the Virasoro constraints. We can see the roots of the above
equation correspond to two different solutions of the string equation of motion
\begin{eqnarray}
 C_1 &=& v ~~~~~  \nonumber\\ &=& \frac{1}{v} ~~~~~  \label{cases}.
\end{eqnarray}
Since for the case $C_1 = \frac{1}{v}$ we would have $\mu = \frac{1}{v}$
then it is a natural choice to work with $C_1 = v$ as detailed in \cite{Ryang:2006yq}.
However, we will discuss qualitative features of both the cases. Also note that
the choice of the constant is supported by demanding forward propagation of
strings in this case, i.e.
\begin{equation}
\dot{t} = \frac{1}{1-v^2}\left(1- \frac{v^2}{\cosh^2\chi}\right) >
0 \ .
\end{equation}
Now, integrating (\ref{chieq}) we can write down the string profile as
\begin{equation}
 y = \pm \frac{(1-v^2)}{\sqrt{1-\omega^2}}\frac{1}{\sqrt{1-\alpha^2}}\tanh^{-1}
 \left[\sqrt{\frac{\cosh^2\chi - \alpha^2}{1 - \alpha^2}}\right] \
 ,
\end{equation}
where $\alpha^2 = \frac{v^2}{1-\omega^2} = \cosh^2\chi_1$ (say) is
a root of the equation $\chi\prime(y) = 0$. As discussed in
\cite{Ryang:2006yq}, \cite{Dai:2014twa} depending on the value of
this root, we can have two different classes of string solutions
for the system.

i) When  $\cosh^2\chi_1 > 1$ the solution can be rewritten in the form
\begin{equation}
y = \pm \frac{(1-v^2)}{\sqrt{v^2+\omega^2-1}} \cos^{-1}
\left(\frac{\sqrt{v^2+\omega^2-1}}{\sqrt{1-\omega^2}\sinh\chi}\right)
\ .
\end{equation}
It is clear from here that $\frac{-\pi}{2\beta}\leq y \leq
\frac{\pi}{2\beta}$, where $\beta=
\frac{\sqrt{v^2+\omega^2-1}}{(1-v^2)}$. These string profiles
correspond to the so called hanging strings as shown in Figure
(\ref{fig:hangingno}). It can be seen that as $\beta(\omega, v)$
increases, the width of the string decreases as range of $y$
becomes smaller.

ii) When  $\cosh^2\chi_1 \leq 1$ the solution can be rearranged in the expression
\begin{equation}
y = \pm \frac{(1-v^2)}{\sqrt{1-v^2-\omega^2}} \sinh^{-1}\left(\frac{\sqrt{1-v^2-\omega^2}}{\sqrt{1-\omega^2}\sinh\chi}\right)
\end{equation}
Here the range of $y$ changes to  $-\infty \leq y \leq \infty$ and
these are the spiky string profiles as shown in Figure
(\ref{fig:spikyno}). The solution is supported in the infinite
range of $y$ which makes it more interesting to study. One can see
from the figures how the string profiles vary under various values
of $(\omega, v)$.

%\begin{figure}
%\centerline{\epsfig{file=hanging_no_kappa.EPS,width=11cm}}
%\vspace{-1.5cm}
%\caption{\small
%one spike of the helical string where the inner and outer circles indicate
%and , respectively.  }
%\label{fR}
%\end{figure}

\begin{figure}
\begin{center}
 \includegraphics[width=0.8\linewidth]{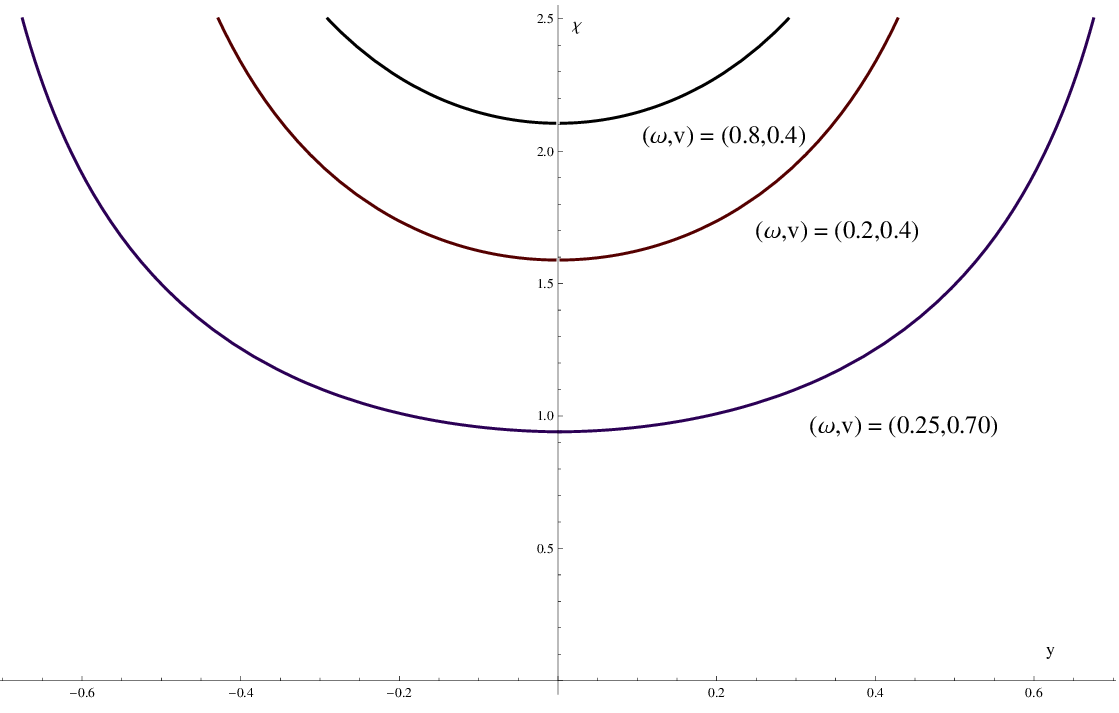}
\caption{Hanging strings for the choice of parameter $C_1 = \frac{1}{v}$ for various $(\omega, v)$}\label{fig:hangingalt}
\end{center}
\end{figure}

However, there is another choice of $C_1 = \frac{1}{v}$, which
leads to the following string equation of motion
\begin{equation}
 \chi\prime(y)= \pm \frac{\sqrt{1-\omega^2}}{(1-v^2)}\tanh\chi \sqrt{\cosh^2\chi - \frac{1}{v^2(1-\omega^2)}} \label{chieq2}
\end{equation}
Since $0 < \omega^2 <1$ and  $0 < v^2 <1$, we can readily show that $\frac{1}{v^2(1-\omega^2)} > 1$
is always true. Then it is evident that we will only get one kind of string solutions, namely
the hanging ones. The string profiles are plotted in Figure (\ref{fig:hangingalt}) for various values of $(\omega, v)$.

\subsection{Open strings in $(AdS_3 \times S^1)_\varkappa$}
To start with we write the total deformed $AdS_3 \times S^3$
metric which has been discussed as a consistent truncation of one
parameter deformed family of $AdS_5 \times S^5$ in
\cite{ABF,Hoare:2014pna}.
\begin{eqnarray}
ds^2 &=& \frac{1}{1-\varkappa^2 \sinh^2\rho}\left[-\cosh^2{\rho}dt^2 + d\rho^2 \right]
+ \sinh^2{\rho}d\phi^2 \nonumber\\
&+& \frac{1}{1+\varkappa^2 \cos^2\theta}\left[\sin^2\theta d
\varphi^2 + d\theta^2\right] + \cos^2\theta d \psi^2
\end {eqnarray}

Now we can see that there is a singularity in the metric for the
value $\rho = \rho_s = \sinh^{-1}\frac{1}{\varkappa}$, which also
manifests itself in the scalar curvature. However we can get rid
of this problem by mapping into a good coordinate system which
describes only the inside of this so-called `singularity surface'.
This transformation has been explicitly discussed in
\cite{Kameyama:2014vma} in the following form,
\begin{equation}
\frac{\cosh\rho}{\sqrt{1-\varkappa^2 \sinh^2\rho}} = \cosh\chi \ ,
\end{equation}
where the range of the $AdS$ radius is mapped from $\rho \in
[0,\sinh^{-1}\frac{1}{\varkappa})$ to $\chi \in [0,\infty)$ and
`zooms' into the region inside the singularity surface. In
\cite{Kameyama:2014vma} it was conjectured that this singualrity
surface might act as a holographic screen inside of which all the
degrees of freedom are confined. In what follows, we will mainly
discuss spiky string solutions which approach $\rho = \rho_s$
instead of $\rho \rightarrow \infty$. Now as it can be shown that
$\theta = \frac{\pi}{2}$ is still a solution of the classical
string equation of motion, we can write the transformed
$AdS_3\times S^1$ part of the deformed metric as,
\begin{equation}
ds^2 = \frac{1}{1+\varkappa^2
\cosh^2\chi}\left[\sinh^2{\chi}d\phi^2 + d\chi^2 \right]-
\cosh^2{\chi}dt^2 + d \varphi^2 \ .
\end{equation}
To study rigidly rotating open strings in the above background, we
again use similar ansatz as in the previous section
\begin{equation} \label{ansatz2}
t=\tau+h_1(y), \quad \chi = \chi(y),\quad \phi = \omega (\tau +
h_2(y)),\quad \varphi= \Omega\, \tau.
\end{equation}
Then from the $t$ and $\phi$ equations of motion we would get,
\begin{equation}
 \partial_y h_1 = \frac{1}{1 - v^2}\left[\frac{A_1}{\cosh^2\chi} - v\right],
 \\\\\\ \ \partial_y h_2 = \frac{1}{1 - v^2}\left[\frac{A_2(1 + \varkappa^2\cosh^2\chi)}{\sinh^2\chi} - v\right].
\end{equation}
As before, the $\chi$ equation supplemented by the boundary condition  $\chi\prime(y)\rightarrow 0$
 as $\chi \rightarrow 0$ we conclude that $A_2 = 0$. Thus the form of the $\chi$ equation becomes
 \begin{equation}
 \chi\prime(y) = \pm \frac{\tanh\chi}{1-v^2}\sqrt{f(\chi)}, \label{profile2}
 \end{equation}
Where $f(\chi)= \varkappa^2 \cosh^4\chi +
(1-\omega^2-A_1^2\varkappa^2)\cosh^2\chi - A_1^2$. Again the
Virasoro constraints give the same results as the previous section
when subtracted. We'll chose $A_1 = v$ for further calculations.
\subsubsection{String profiles and the effect of $\varkappa$}
\begin{figure}
\begin{center}
 \includegraphics[width=0.85\linewidth]{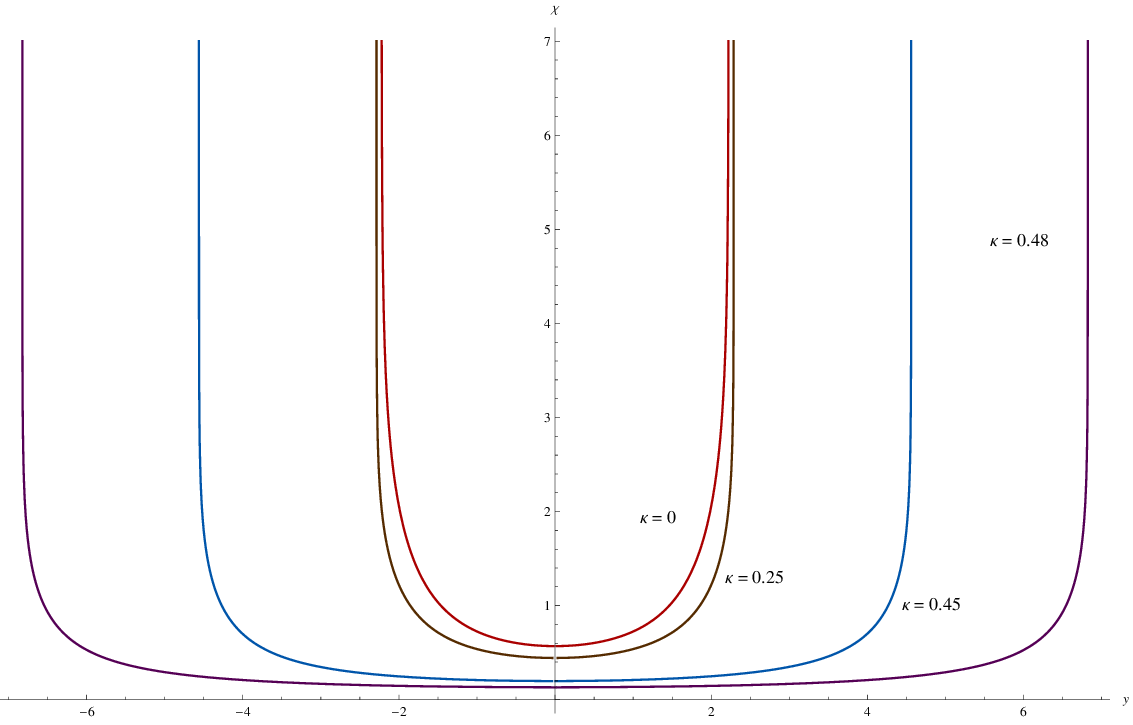}
\caption{Hanging string profiles for various values of $\varkappa$ with fixed $(\omega, v) = (0.8,0.7)$.
Notice the flattening of the profile as $\varkappa$ is increased.}\label{fig:hangingyes}
\end{center}
\end{figure}

\begin{figure}[!]
\begin{center}
 \includegraphics[width=0.85\linewidth]{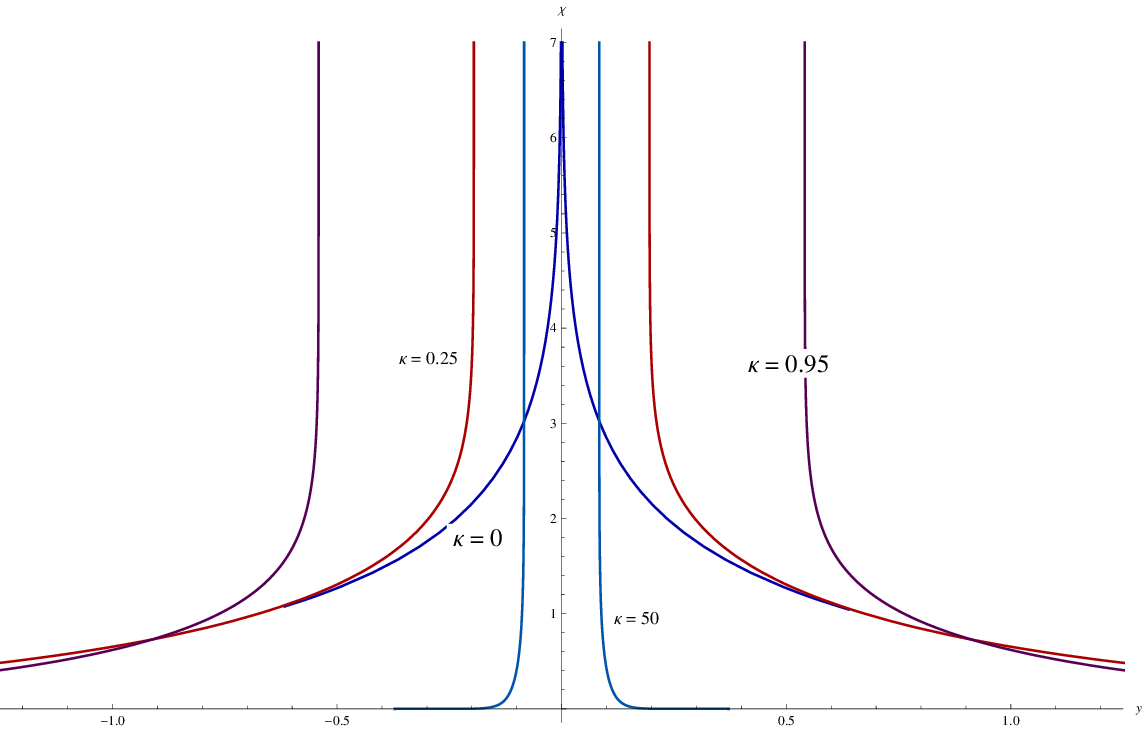}
\caption{Spiky string profiles for various values of $\varkappa$ for fixed $(\omega, v) = (0.2,0.4)$.}\label{fig:spikyyes}
\end{center}
\end{figure}

To find the rotating string profile we have to integrate equation (\ref{profile2})
between some appropriate range. We note that since $f(\chi)$ is  quadratic function
in $\cosh^2\chi$, for the parameter space $(\omega, v)\in(0,1)$ and $\varkappa \in [0,\infty)$
we can always find two real roots of the function,
\begin{equation}
\cosh^2\chi_{\pm} = \frac{-1+\omega^2+v^2\varkappa^2 \pm
\sqrt{(-1+\omega^2+v^2\varkappa^2)^2 +
4v^2\varkappa^2}}{2\varkappa^2} \ ,
\end{equation}
where $\chi_{+} > \chi_{-}$. The string profile can then be
written in the integral form
\begin{equation}
y = \pm \int \frac{(1-v^2)}{\varkappa \tanh\chi \sqrt{(\cosh^2\chi
- \cosh^2\chi_{+})(\cosh^2\chi - \cosh^2\chi_{-})}} d\chi \ .
\end{equation}
Now the string solutions can exist only in the cases $f(\chi)
\geqslant 0$. This can only be possible if either $\chi \geqslant
\chi_{+}$ or $\chi \leqslant \chi_{-}$. It can be cleary seen that
the $\varkappa = 0$ string profiles are not smoothly connencted to
the finite $\varkappa$ profiles. However it is notable that for a
fixed $\varkappa$ the above expression depends on the parameters
$(\omega,v)$ only. The solutions can be classified by comparison
of $\chi_{+}$ and $\chi_{-}$. Again a simple analysis shows that
there are two kinds of string solution possible, the hanging
strings and the spiky strings. Spiky strings extend all the way
upto the asymptotic infinity and are the solutions we are
interested in.

i) Firstly, it can be clearly seen that $\cosh^2\chi_{-}$ is always negative for
$(\omega, v, \varkappa)\in \mathcal{R}$. The hanging string profiles exist for
the region where $\cosh^2\chi_{+} \geqslant 1$. For example using $(\omega, v) = (0.8,0.7) $
this inequality is violated at $\varkappa \backsimeq 0.503$. As $\varkappa$ increases the
hanging strings are more and more flattened as the range of $y$ increases. This case is
depicted in Figure (\ref{fig:hangingyes}).

ii) The second case is when $\cosh^2\chi_{+} < 1$, which give rise to the spiky strings.
Note that in this region the value of the root remains almost constant for the changing
values of $\varkappa$. It can be seen from Figure (\ref{fig:spikyyes}) that as $\varkappa$
increases the spiky `cusp' nature of the strings are destroyed. For large values of the deformation
parameter, the strings simply become parallel to the $\chi$ axis.
\subsubsection{Conserved charges and dispersion relation}
We can construct the following conserved charges for the rotating
string solutions looking at the symmetries of the background. They
are
\begin{eqnarray}
E &=& -\int\frac{\partial \mathcal{L}}{\partial \dot{t}}d\sigma
=
\frac{2\hat{T}}{(1-v^2)}
\int\left(\cosh^2\chi-v^2 \right)d\sigma \ , \nonumber\\
S &=& \int\frac{\partial \mathcal{L}}{\partial\dot{\phi}}~d\sigma
= 2\hat{T}\frac{\omega}{(1-v^2)}\int\frac{\sinh^2\chi}{1+\varkappa^2\cosh^2\chi}~d\sigma \ , \nonumber\\
J &=& \int\frac{\partial
\mathcal{L}}{\partial\dot{\varphi}}~d\sigma = 2\hat{T} \Omega\int
d\sigma \ ,
\end{eqnarray}
where $\hat{T} = \frac{\sqrt{\lambda}}{2\pi}\sqrt{1 +
\varkappa^2}$. It can be seen that the above charges satisfy a
relation
\begin{equation}
E - \frac{J}{\Omega} = \frac{S}{\omega}+ \mathcal{K} \ ,
\end{equation}
where $\mathcal{K}$ acts like a correction term to the dispersion
realtion. Note that, all the charges themselves are dependent on
$\varkappa$. The expression for $\mathcal{K}$ given by
\begin{equation}
\mathcal{K} = \frac{2\hat{T}}{(1-v^2)}\int\frac{\varkappa^2\sinh^2\chi \cosh^2\chi}
{1+\varkappa^2\cosh^2\chi}~d\sigma.
\end{equation}
It can be explicitly checked that when we take the $\varkappa = 0$ the correction piece
goes to zero also and the dispersion relation reduces to the exact one
$E - \frac{J}{\Omega} = \frac{S}{\omega}$ as given in \cite{Ryang:2006yq}.
%\begin{figure}
%\begin{center}
% \includegraphics[width=0.8\linewidth]{E_numerical.eps}
%\caption{Energy of the rotating strings plotted against $\omega$ for different
%                values of $\varkappa$ with $v = 0.4$. For very large values of deformation
%                we can see the Energy becomes more
%                or less indifferent to the change in $\omega$
%                The cutoff is implemented at $\chi_{\Lambda} = 50$.}\label{fig:Energyopen}
%\end{center}
%\end{figure}
%\begin{figure}
%\begin{center}
% \includegraphics[width=0.8\linewidth]{J_numerical.eps}
%\caption{Angular momenta of the rotating strings
%                coming from the sphere part plotted against $\omega$ for different
%                values of $\varkappa$ with $v = 0.4$. For very large values of deformation
%                we can see it exactly behaves like the Energy.
%                The cutoff is implemented at $\chi_{\Lambda} = 50$.}\label{fig:Angmomopen}
%\end{center}
%\end{figure}
\begin{figure}
\begin{center}
 \includegraphics[width=0.7\linewidth]{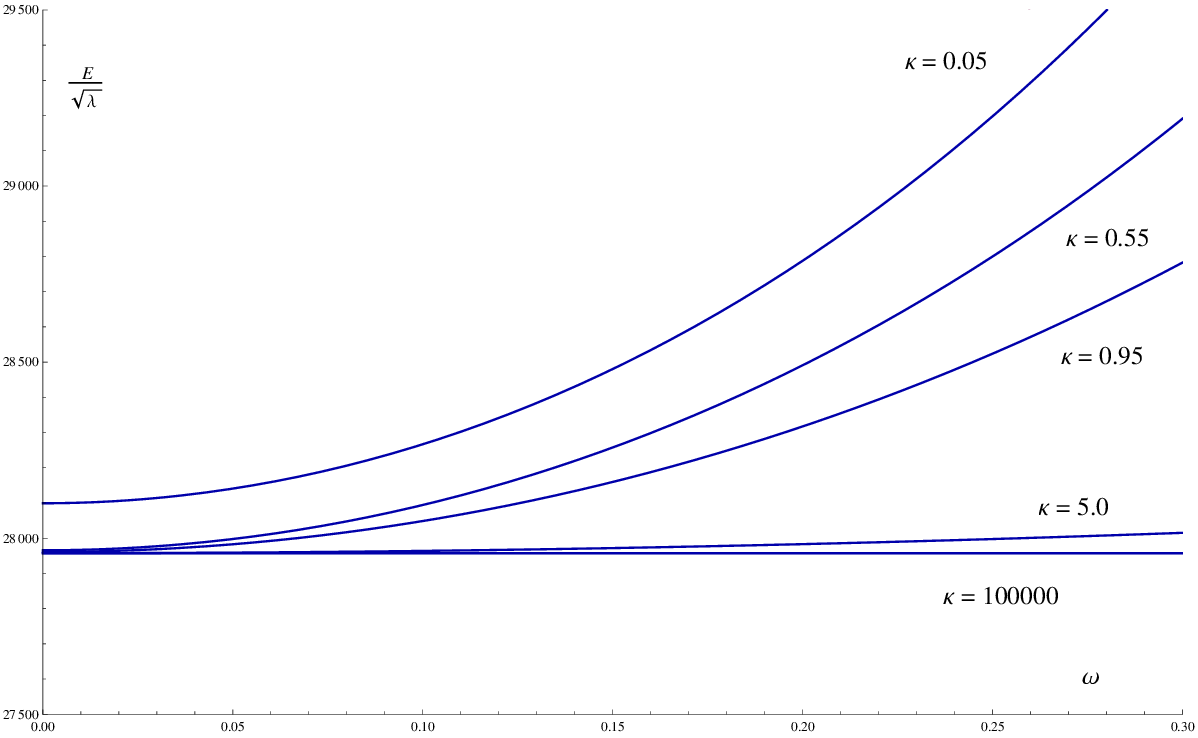}
 \caption{Energy of the rotating strings plotted against $\omega$ for different
                values of $\varkappa$. For very large values of deformation
                we can see the Energy becomes more
                or less indifferent to the change in $\omega$
                The cutoff is implemented at $\chi = 50$.}\label{fig:Energyopen}
\end{center}
\end{figure}
\begin{figure}
\begin{center}
 \includegraphics[width=0.7\linewidth]{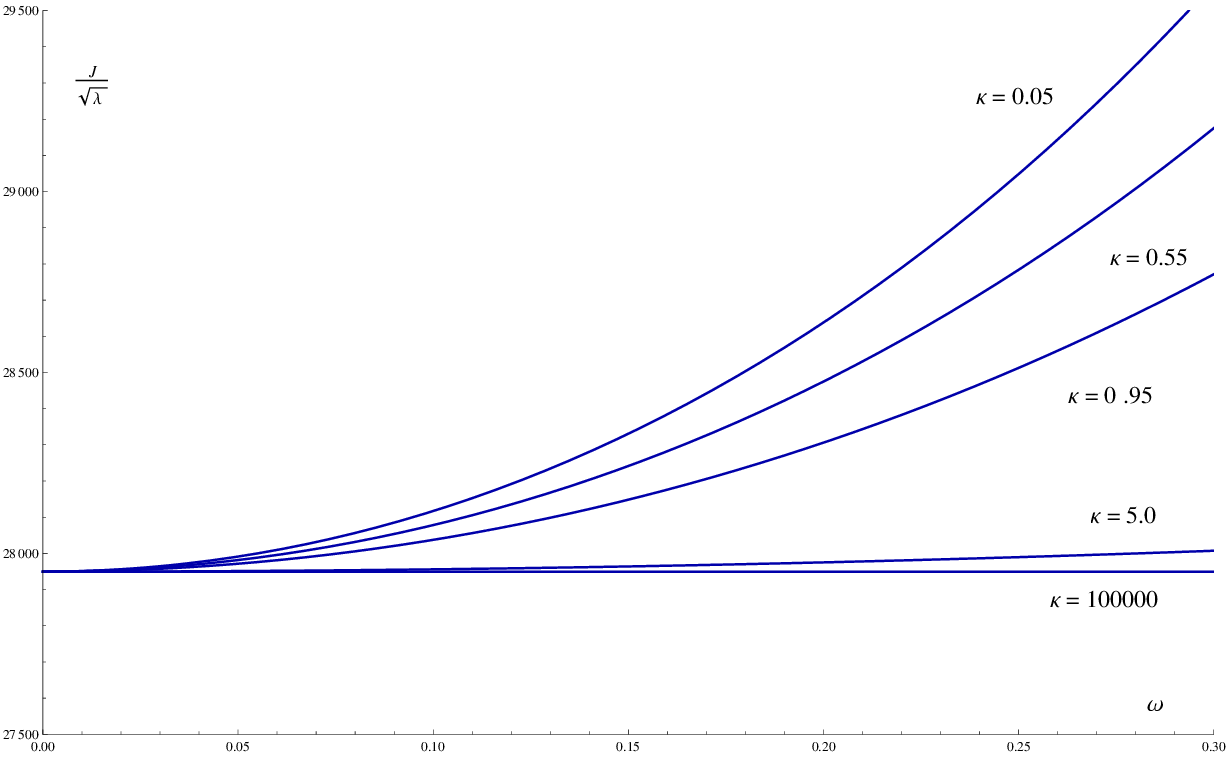}
\caption{Angular momenta of the rotating strings
                coming from the sphere part plotted against $\omega$ for different
                values of $\varkappa$. For very large values of deformation
                we can see it exactly behaves like the Energy.
                The cutoff is implemented at $\chi = 50$.}\label{fig:Angmomopen}
\end{center}
\end{figure}
Now as the charges are  integrated upto $\chi = \infty$, they
diverge. Since the integrals involved are really complex, we can't
obtain regularized expressions for $E$,$S$ and $J$ analytically.
Instead we focus on the nature of the charges as they change with
the value of winding number $\omega$. For a numerical computation,
we impose a cutoff at $\chi = 50$ and plot the expressions in the
Figures (\ref{fig:Energyopen}) and (\ref{fig:Angmomopen}). From
the plot it can be seen that while $E$ and $J$ have large values
for small winding number $\omega$, their difference remains small
and finite.
\begin{figure}
        \centering
        \begin{subfigure}[b]{0.5\textwidth}
                \includegraphics[width=\textwidth]{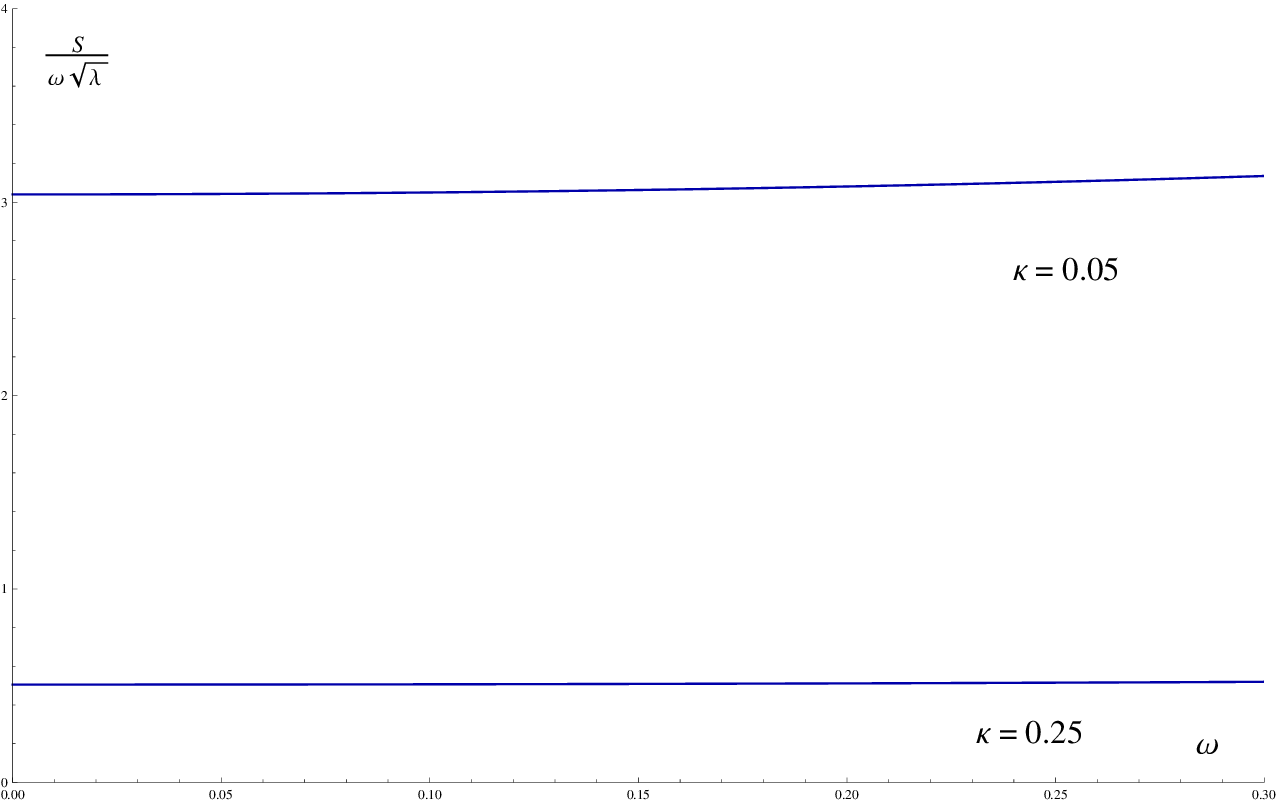}
                \caption{}
                \label{fig:Sopen}
        \end{subfigure}%
        ~ %add desired spacing between images, e. g. ~, \quad, \qquad, \hfill etc.
        \begin{subfigure}[b]{0.5\textwidth}
                \includegraphics[width=\textwidth]{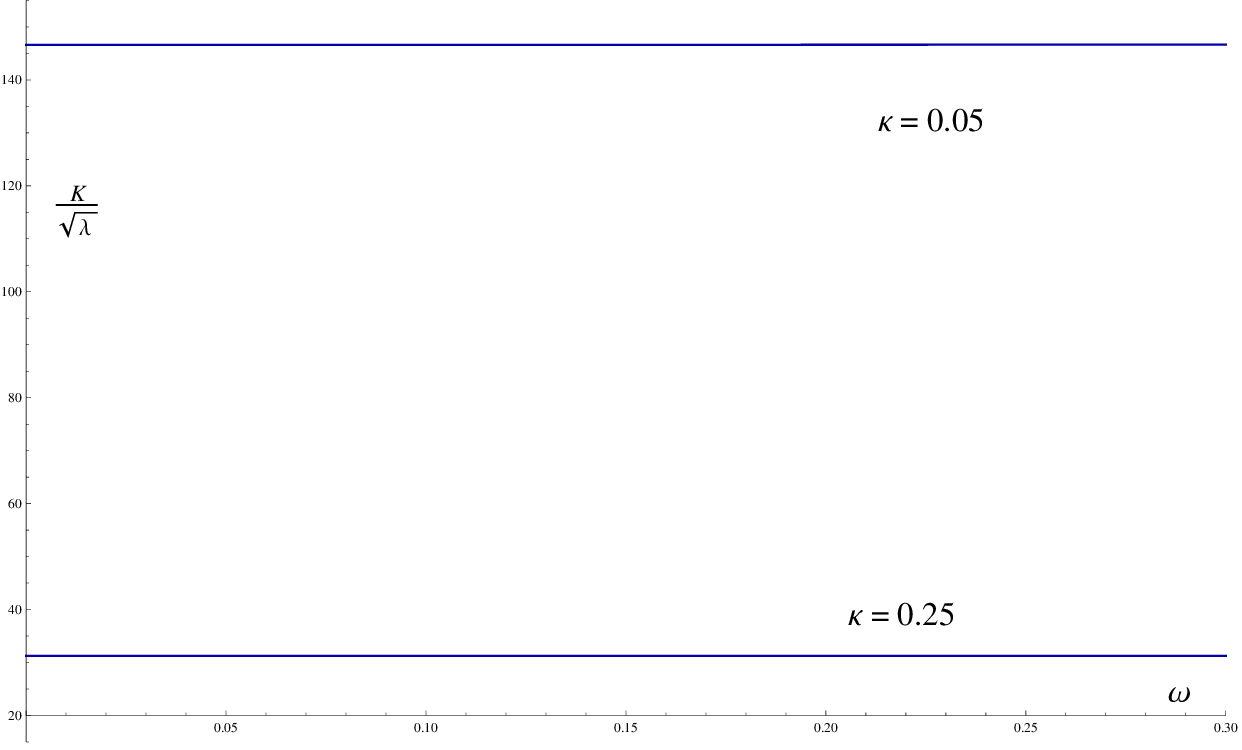}
                \caption{}
                \label{fig:Kopen}
        \end{subfigure}

        \caption{a)\,Spin $\frac{S}{\omega}$ of the rotating strings plotted against $\omega$ for different
                values of $\varkappa$ with $v = 0.4$.
              The cutoff is again at $\chi_{\Lambda} = 50$. b)\,$\mathcal{K}$ of the spiky strings plotted against $\omega$ for different
                values of $\varkappa$ with $v = 0.4$.
                The integration cutoff is at $\chi_{\Lambda} = 50$.}\label{fig:K_and_S}
\end{figure}

Similarly we also show the behaviour of $\frac{S}{\omega}$ and
$\mathcal{K}$ for various values of the winding number in Figures
(\ref{fig:Sopen}) and (\ref{fig:Kopen}). It is worth noting that
both the expressions are extremely sensitive to $\varkappa$ for
fixed values of $v$. For a fixed $\varkappa$ these expressions
remain almost constant with changing $\omega$.
\section{Closed strings and spikes in one parameter deformed $AdS$}
It is worth mentioning, that closed strings in the $\varkappa$
deformed $AdS$ have been studied in \cite{frolov,
Kameyama:2014vma} in the GKP limit. The long string limit of the
GKP solution (extending to the apparent boundary) in this case has
been shown to be quite complex. One interesting feature is that
the long string dispersion relation does not reduce to the usual
scaling $E - S \sim \log S$ for undeformed $AdS$, even when
$\varkappa \rightarrow 0$. Also it has been shown that the
classical spinning string can not be stretched beyond the
singularity surface all the way to the real boundary. This prompts
us to look for symmetric closed spiky string solutions first
discussed in \cite{Kruczenski:2004wg} for $AdS_3$ and see if they
have the same kind of dynamics as the GKP ones. We will review the
Nambu-Goto spiky strings in the simple $AdS$ background and try to
compare them with deformed $AdS$ calculations.
\subsection{Review of Nambu-Goto spiky strings in $AdS$}
Let us consider Nambu-Goto strings moving in the $AdS_3$ space parameterized by the
metric
\begin{equation}
ds^2 = -\cosh^2{\chi}dt^2 + d\chi^2 + \sinh^2{\chi}d\phi^2 \ ,
\end{equation}
with the following embedding ansatz,
\begin{equation}
 t= \tau , \ \ \phi = \omega\tau + \sigma, \ \ \chi = \chi(\sigma)
 \ .
\label{soc}
\end{equation}
The Nambu-Goto action for this fundamental string as
\begin{equation}
 S = - \frac{\sqrt{\lambda}}{2\pi} \int \sqrt{-\dot{X}^2 {X'}^2 +
 (\dot{X}X')^2} \ .
\label{NGaction}
\end{equation}
The dots represent derivative w.r.t $\tau$ and primes represent
derivatives w.r.t $\sigma$ for the notation of this section. Now
following the equations of motion for $t$ and $\phi$ we can write
\begin{eqnarray}
 \frac{\sinh^2\chi\cosh^2\chi}{ \sqrt{{\chi'}^2(\cosh^2\chi-\omega^2\sinh^2\chi)+\sinh^2\chi\cosh^2\chi}} = C
 \nonumber \\
\Rightarrow \ \
\chi\prime = \frac{1}{2}\frac{\sinh2\chi}{\sinh2\chi_0}\frac{\sqrt{\sinh^22\chi-\sinh^22\chi_0}}
{\sqrt{\cosh^2\chi-\omega^2\sinh^2\chi}}
\label{rhoeq}
\end{eqnarray}
Here the integration constant is taken as $C^2 =
\frac{1}{4}\sinh^2 2\chi_0$. Now the r.h.s of the above equation
blows up at  $\chi_1=\coth^{-1}\,\omega$, which gives rise to the
spiky string solutions with the spike height $\chi_1$. The
`valleys' of the string profile occurs at the zeroes of $\chi'$,
i.e at $\chi = \chi_0$. Integrating the above equation, we can get
the string profile for our rigidly rotating ansatz as the
following,
\begin{equation}
\sigma = \frac{\sinh2\chi_0}{\sqrt{2}\sqrt{\alpha_0+\alpha_1}\sinh\chi_1}
  \left\{\Pi(\frac{\alpha_1-\alpha_0}{\alpha_1-1},\Theta,\mathcal{Q})-
  \Pi(\frac{\alpha_1-\alpha_0}{\alpha_1+1},\Theta,\mathcal{Q})
  \right\} \ ,
\label{alpha}
\end{equation}
where $\Pi$ is the standard elliptic integral of the third kind and
\begin{eqnarray}
 \mathcal{Q} &=& \frac{\alpha_1-\alpha_0}{\alpha_1+\alpha_0} \ , \label{pdef}\\
 \sin\Theta &=&  \sqrt{\frac{\alpha_1-\alpha}{\alpha_1-\alpha_0}} \ , \label{alphadef}
\end{eqnarray}
with $\alpha = \cosh2\chi, \ \ \ \alpha_0=\cosh 2\chi_0, \ \ \
\alpha_1=\cosh2\chi_1$. The shape above, supplemented by the
closedness condition on the string i.e. $\Delta\phi =
\frac{2\pi}{2n}, \ \ n\in\mathbb{Z}$, gives the total string
profile. Here $n$ is the total number of spikes formed.
\begin{figure}
\begin{center}
 \includegraphics[width=0.6\linewidth]{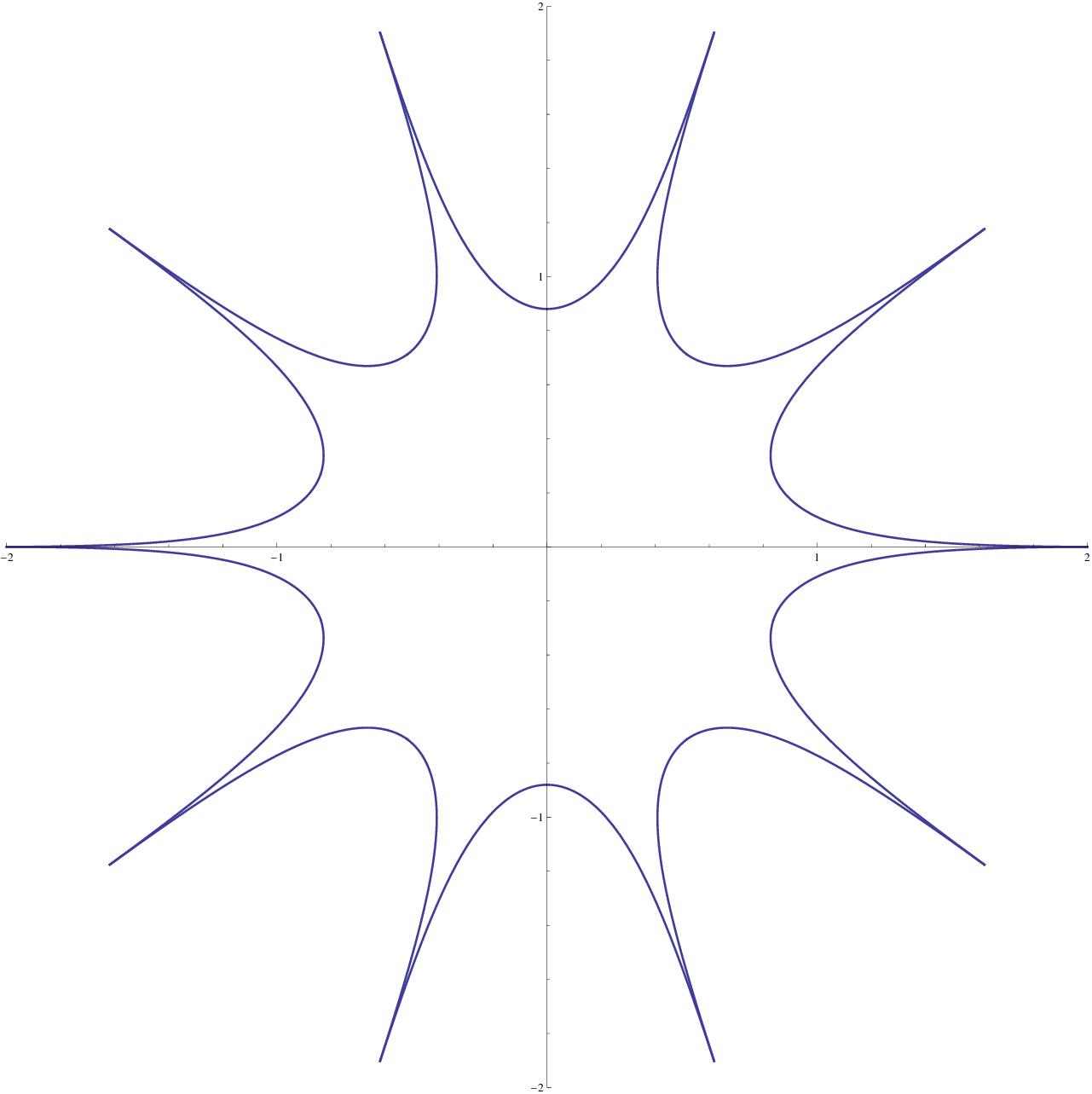}
 \caption{Spiky strings in AdS with 10 spikes, parameters $\chi_1 = 2.0$ and $\chi_0 = 0.88$   .}\label{fig:Spikes10}
\end{center}
\end{figure}
The angle between the valleys and the spikes is given by,
\begin{equation}
\Delta \phi =  2 \int_{\chi_0}^{\chi_1}\frac{\sinh2\chi_0}{\sinh2\chi}
 \frac{\sqrt{\cosh^2\chi-\omega^2\sinh^2\chi}}{\sqrt{\sinh^22\chi-\sinh^22\chi_0}}\, d\chi
\end{equation}
Again, the change of variables as before gives the solution
\begin{equation}
\Delta \phi = \frac{\sinh2\!\chi_0}{\sqrt{2}\sinh\chi_1}\frac{1}{\sqrt{\alpha_0+\alpha_1}}
 \left\{\Pi(\frac{\alpha_1-\alpha_0}{\alpha_1-1},\frac{\pi}{2},\mathcal{Q})-\Pi(\frac{\alpha_1-\alpha_0}{\alpha_1+1},\frac{\pi}{2},\mathcal{Q}) \right\}.
\end{equation}
Clearly $\Delta \phi$ is dependent on the parameters $\chi_1$ and
$\chi_0$. Now to implement the closed string condition we fix the
height of the tip of spike by adjusting $\omega$. Following
\cite{Kruczenski:2004wg} we fix the spike height at $\chi_1 = 2$
and $n=10$. Then from the closeness condition, solving for
$\chi_0$ we get $\chi_0 \simeq 0.88$. Using these values
(\ref{alpha}) gives the plot of one spike, which we can rotate and
reflect to plot the others. Note that the spikes will reach the
$AdS$ boundary if we make $\chi_1 \rightarrow \infty$ i.e. $\omega
\rightarrow 1$. Spiky strings in this `long' string limit has been
discussed in \cite{Kruczenski:2004wg}. The total plot of the
spikes is shown in the figure (\ref{fig:Spikes10}).
\subsection{Nambu-Goto spiky strings in $(AdS)_\varkappa$}
We start with the one parameter deformed $AdS_3$ background, and
write it using the coordinate transformations discussed earlier.
The metric as used in \cite{Kameyama:2014vma} is
\begin{equation}
ds^2 =
-\cosh^2{\chi}dt^2+\frac{1}{1+\varkappa^2\cosh^2{\chi}}(d\chi^2+\sinh^2{\chi}d\phi^2)
\ .
\end{equation}
Here we use the same embedding ansatz as the simple $AdS$ case i.e ,
\begin{equation}
t=\tau,\ \ \phi=\omega\tau+\sigma , \ \ \chi=\chi(\sigma) \ .
\label{ansatzkappa}
\end{equation}
Now as before, we will use the Nambu-Goto action to write the equations of motion of
the fundamental string. Simply following the equations of motion for $t$ and $\phi$ we can write,
\begin{equation}
\sinh^2{\chi}\cosh^2{\chi}\\ =
C\sqrt{({\chi'}^2\cosh^2{\chi}+\sinh^2{\chi}\cosh^2{\chi})
(1+\varkappa^2\cosh^2{\chi})-\omega^2{\chi'}^2\sinh^2{\chi}} \ ,
\end{equation}
where $C$ is a integration constant. Now this can be shown to
satisfy the equation of motion for $\chi$, which we write as,
\begin{equation}
\chi'=\frac{1}{2}\frac{\sinh{2\chi}}{\sinh{2\chi_0}}\frac{\sqrt{\sinh^2{2\chi}-
(1+\varkappa^2\cosh^2{\chi})\sinh^2{2\chi_0}}}{\sqrt{(1+\varkappa^2\cosh^2{\chi})
\cosh^2{\chi}-\omega^2\sinh^2{\chi}}} \ . \\
\end{equation}
Here the integration constant is taken as before
$C^2=\frac{1}{4}\sinh^2{2\chi_0}$. Now the above equation can be
arranged to write the string profile as
  \begin{equation}
  \sigma =  \varkappa\,\sinh 2\chi_0 \int \frac{1}{\sinh 2\chi}
  \frac{\sqrt{(\cosh 2\chi - \cosh 2\chi_1^{+})(\cosh 2\chi - \cosh 2\chi_1^{-})}}
  {\sqrt{(\cosh 2\chi - \cosh 2\chi_0^{+})(\cosh 2\chi - \cosh 2\chi_0^{-})}}\, d\chi  \
  ,
  \label{chieqn3}
  \end{equation}
where one can notice that the $\chi\prime$ blows up for two values
of $\chi$, they are
\begin{equation}
\chi^{\pm}_1=\frac{1}{2}\cosh^{-1}{\frac{(\omega^2-\varkappa^2-1)\pm\sqrt{1-2(1+2\varkappa^2)
\omega^2+\omega^4}}{\varkappa^2}} \ , \label{chiplus}
\end{equation}
so that we can say $\chi_1$ represents the analogue to spike
height of the corresponding string profile in this geometry. Note
that only $\chi_1^{-}$ reduces to the original spike height in
$AdS$ when we take the $\varkappa \rightarrow 0$ limit in the
proper way. The other root will not be interesting here.

Obviously, the ``valleys'' of the string profile now occurs at $\chi^{\pm}_0$
which we get from the following expression:
\begin{equation}
\chi^{\pm}_0=\frac{1}{2}\cosh^{-1}\frac{\varkappa^2\sinh^2{2\chi_0}\pm\sqrt{\varkappa^4\sinh^4{2\chi_0}
+(8\varkappa^2+16)\sinh^2{2\chi_0}+16}}{4}
\end{equation}
Now integrating the equation (\ref{chieqn3}) we can write the
total string profile as
\begin{eqnarray}
&&\frac{ 2\sigma}{\varkappa\sinh 2\chi_0} =
\frac{2(\cosh{2p}-\cosh{2r})}
{\cosh^2{2r}-1}\sqrt{\frac{\cosh{2q}-\cosh{2r}}{\cosh{2p}-\cosh{2s}}}
\,\mathbb{F}\left(\beta,\nu \right) \  \nonumber \\
&+& \frac{(\cosh{2r}-\cosh{2s})}{\sqrt{(\cosh{2p}-\cosh{2s})(\cosh{2q}-\cosh{2r})}}
\Biggr[ \frac{(\cosh{2p}-1)(\cosh{2q}-1)}{(\cosh{2r}-1)(\cosh{2s}-1)}
\,\Pi\left(\gamma_{-},\beta,\nu \right) \  \nonumber\\
&-& \frac{(1+\cosh{2p})(1+\cosh{2q})}{(1+\cosh{2r})(1+\cosh{2s})}
\,\Pi \left(\gamma_{+},\beta,\nu \right) \Biggr] \ .
\end{eqnarray}
Here $\mathbb{F}$ and $\Pi$ are the usual incomplete Elliptic integrals of the second and third kind.
We have also introduced the notation  $p=\chi^+_1$ ,  $q=\chi^-_1$ ,  $r=\chi^-_0$ ,  $s=\chi^+_0$
in the above equations for simplicity.
Others symbols are defined as follows,
\begin{eqnarray}
 \beta &=& \sin^{-1}{\sqrt{
\frac{(\cosh{2q}-\cosh{2r})(\cosh{2s}-\cosh{2\chi})}
{(\cosh{2q}-\cosh{2s})(\cosh{2r}-\cosh{2\chi})}}} \ , \label{betadef}\\
 \nu &=& \frac{(\cosh{2p}-\cosh{2r})(\cosh{2q}-\cosh{2s})}
{(\cosh{2q}-\cosh{2r})(\cosh{2p}-\cosh{2s})} \ , \label{deltadef}\\
\gamma_{\pm} &=& \frac{(\cosh{2r} \pm
1)(\cosh{2q}-\cosh{2s})}{(\cosh{2q}-\cosh{2r})(\cosh{2s} \pm 1)} \
.\label{gammadef}
\end{eqnarray}
Now we find the corresponding angle between valley and tip of the spike by choosing proper
limits of the integral (\ref{chieqn3}) since we can see that $\cosh 2r$ is always negative
and $\cosh 2p > \cosh 2q > \cosh 2s $. With suitable choice of limits we evaluate the
integral as:
\begin{eqnarray}
&&\frac{2\Delta\phi}{\varkappa\sinh 2\chi_0} =
\frac{2(\cosh{2p}-\cosh{2r})}{\cosh^2{2r}-1}
\sqrt{\frac{\cosh{2q}-\cosh{2r}}{\cosh{2p}-\cosh{2s}}}
\,\mathbb{F}\left(\frac{\pi}{2},\nu \right) \  \nonumber \\
&+& \frac{(\cosh{2r}-\cosh{2s})}{\sqrt{(\cosh{2p}-\cosh{2s})(\cosh{2q}-\cosh{2r})}}
\Biggr[ \frac{(\cosh{2p}-1)(\cosh{2q}-1)}{(\cosh{2r}-1)(\cosh{2s}-1)}
\,\Pi\left(\gamma_{-},\frac{\pi}{2},\nu \right) \  \nonumber\\
&-& \frac{(1+\cosh{2p})(1+\cosh{2q})}{(1+\cosh{2r})(1+\cosh{2s})}
\,\Pi \left(\gamma_{+},\frac{\pi}{2},\nu \right) \Biggr] .
\end{eqnarray}
%\begin{equation}
%\Delta \phi= 2 \int_{\chi_0}^{\chi_1} \frac{\sinh{2\chi_0}}{\sinh{2\chi}}
%\frac{\sqrt{(1+\varkappa^2\cosh^2{\chi})\cosh^2{\chi}-\omega^2\sinh^2{\chi}}}{\sqrt{\sinh^2{2\chi}-(1+\varkappa^2\cosh^2{\chi})\sinh^2{2\chi_0}}} d\chi
%\end{equation}
Here one can see that the expression has been integrated from $\chi = \chi_0^{+}$ to $\chi = \chi_1^{-}$
following our original notation.
Since the values where the tip and `valley'
of the spikes are supposed to occur are complex functions of the
parameters, the presence of desired solutions are also likely
to be constrained by the parameter space of $(\varkappa, \omega, \chi_0)$.
As the parameter space here is substantially bigger than the case in the previous section,
we can get different classes of
string profiles which can be smoothly deformed into one another by simply tweaking
the parameters.
\subsubsection{String profiles}
\begin{figure}
        \centering
        \begin{subfigure}[b]{0.5\textwidth}
                \includegraphics[width=\textwidth]{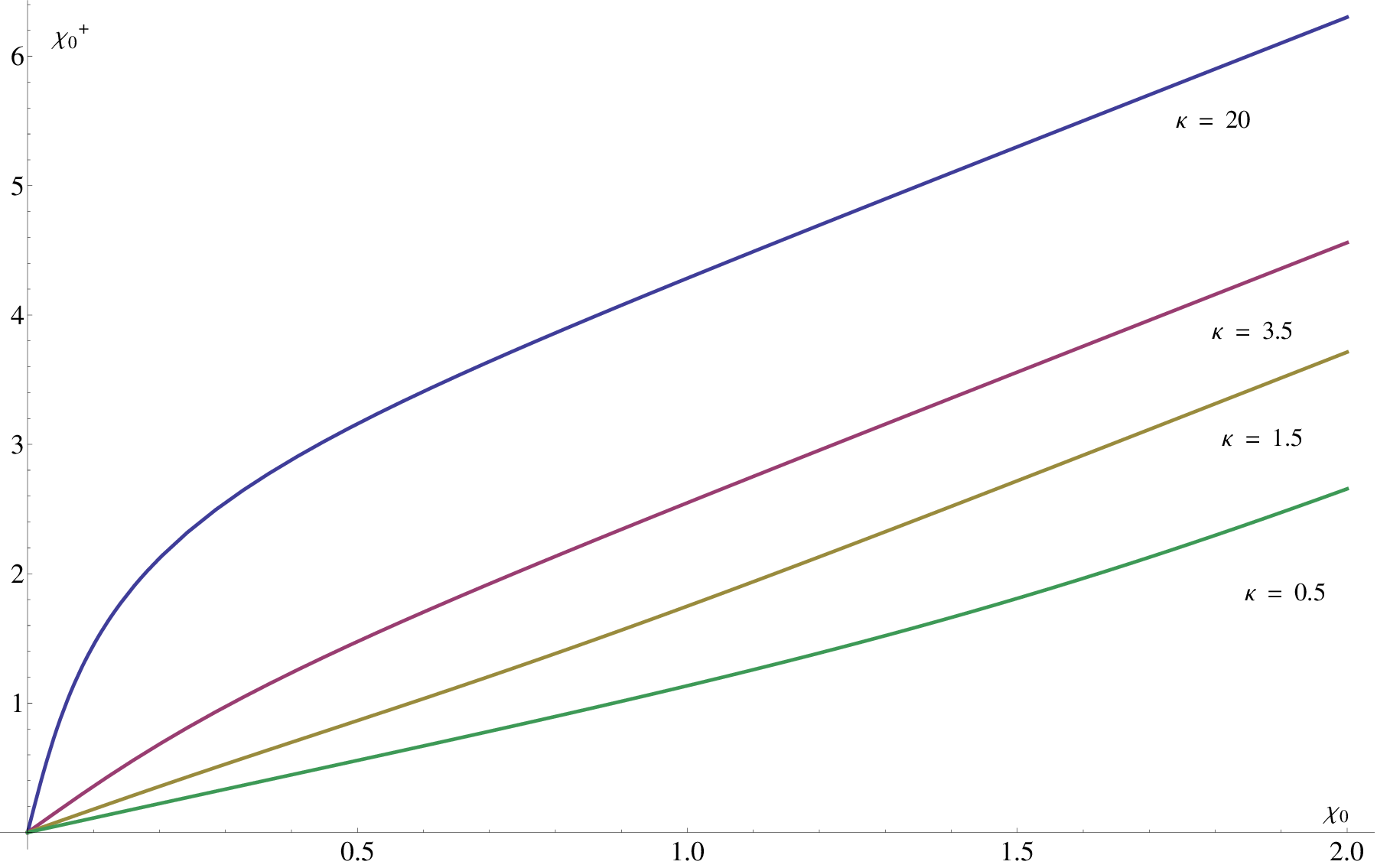}
                \caption{}
                \label{fig:Rootlower}
        \end{subfigure}%
        ~ %add desired spacing between images, e. g. ~, \quad, \qquad, \hfill etc.
        \begin{subfigure}[b]{0.5\textwidth}
                \includegraphics[width=\textwidth]{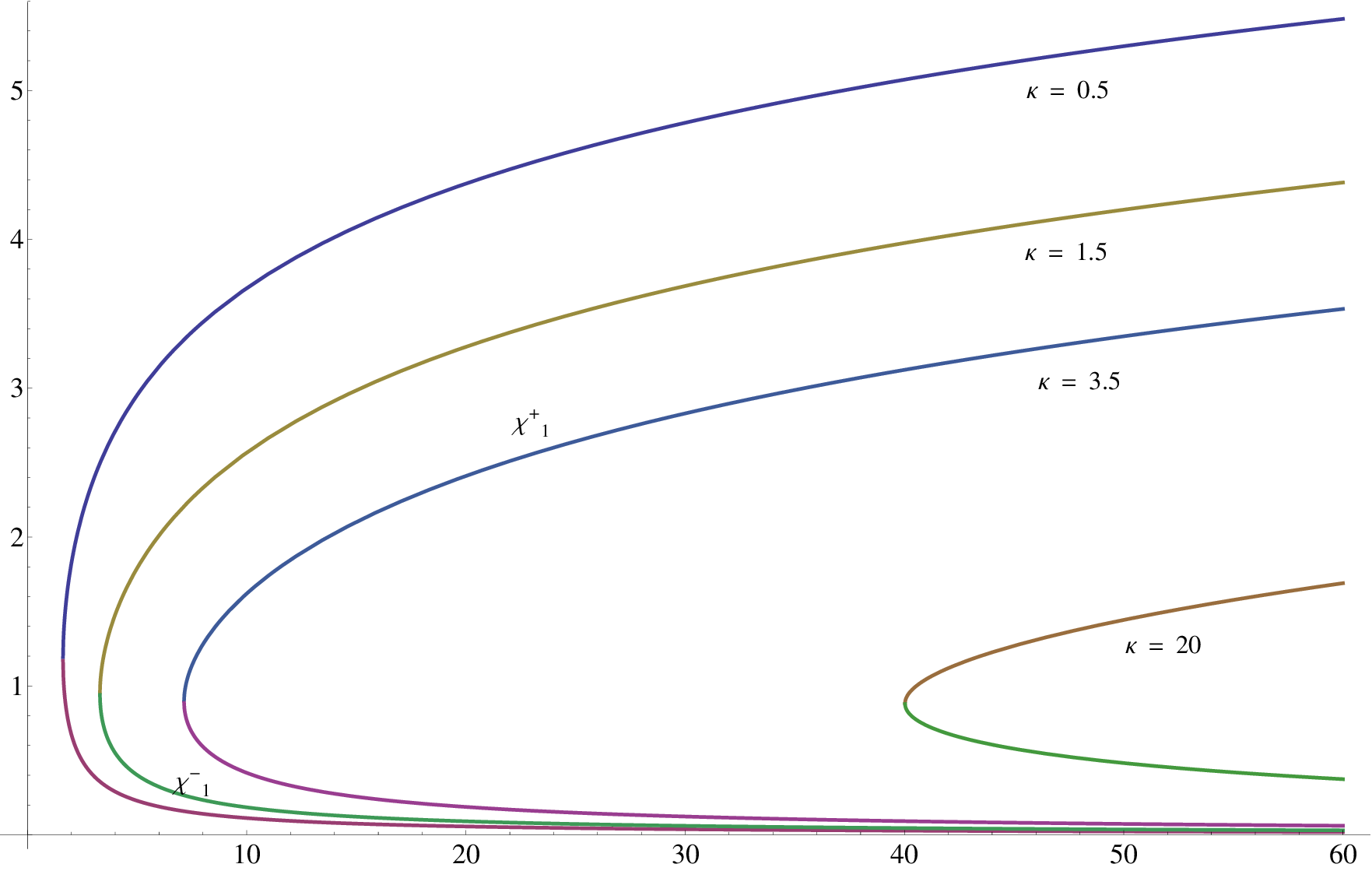}
                \caption{}
                \label{fig:Rootupper}
        \end{subfigure}

        \caption{a)\,$\chi_0^{+}$ plotted for different values of
        $\varkappa$ against increasing $\chi_0$. b)\,$\chi_1^{\pm}$ plotted for different values of
        $\varkappa$ against increasing $\omega$.}\label{fig:roots}
\end{figure}

Now we want to show the spiky profiles graphically and classify
them. First, let us start examining the structure of the positive
roots  $\chi = \chi_1^{\pm},\chi_0^{+}$. We plot the roots in the
figure (\ref{fig:roots}) for different values of $(\varkappa,
\omega, \chi_0)$. It is notable that (Fig. \ref{fig:Rootupper})
with increasing $\omega$, the plots for $\chi_1^{\pm}$ shifts
along the $\omega$ axis. It is also important that the value of
upper limit of integration $\chi_1^{-}$ approaches zero
asymptotically. To make the spike height very large it seems we
have to take $\omega \rightarrow 1$ and $\varkappa\rightarrow 0$
simultaneously, which is obviously the limit for the original $AdS$ 
solution. One can also see that for small values of
$\varkappa$ the lower limit of integration $\chi_0^{+}$ varies
almost linearly with the parameter $\chi_0$. As $\chi_0
\rightarrow 0$ we can see
 $\chi_0^{+} \rightarrow 0$ also. These facts will be relevant when we will
 discuss the dynamics of the spikes in the long string limit. This more or less constrains the
values of the parameter space where our desired solutions can form.

\begin{figure}[t!]
        \centering
        \begin{subfigure}[b]{0.5\textwidth}
                \includegraphics[width=\textwidth]{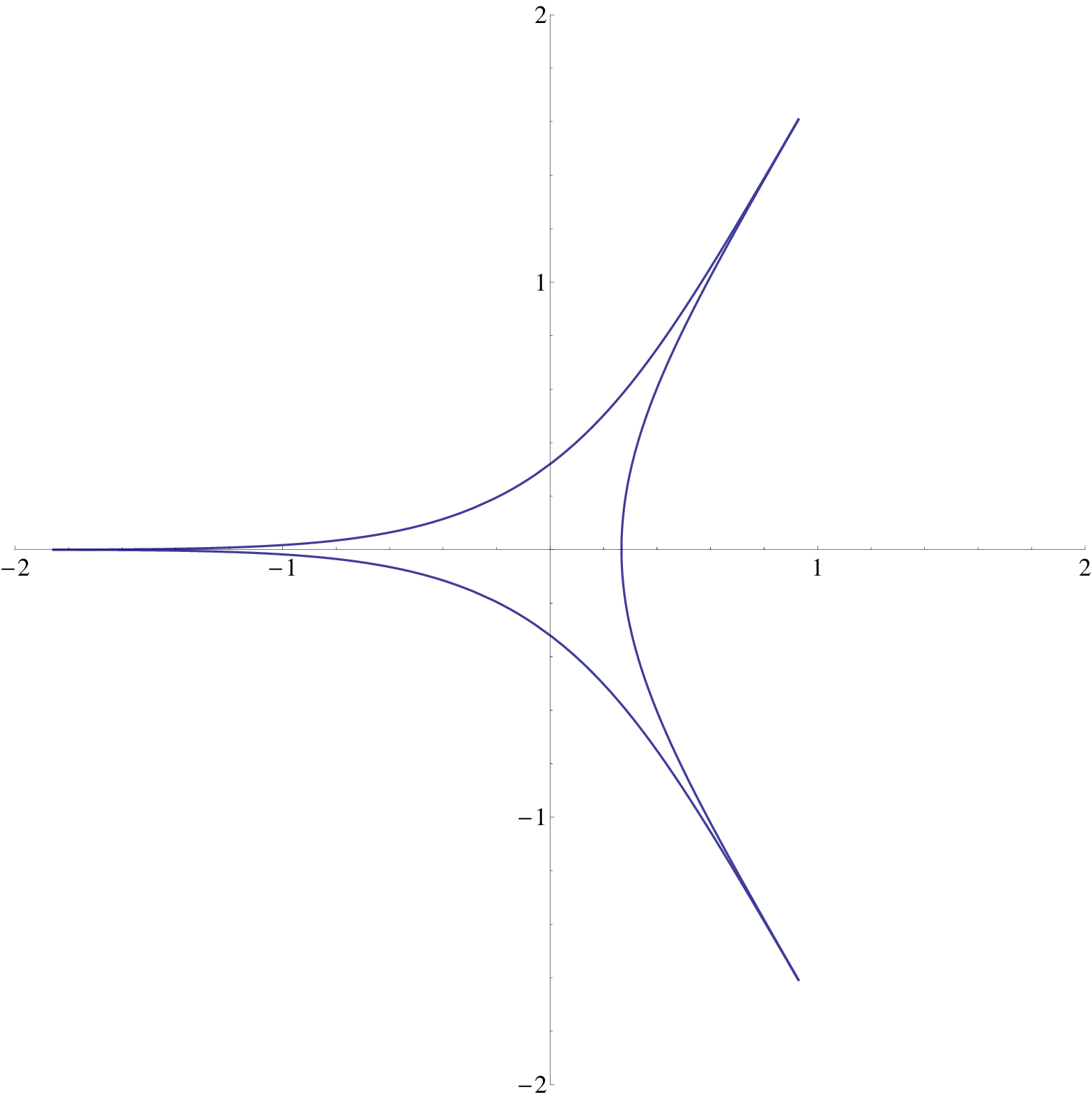}
                \caption{}
                \label{fig:k=0.5}
        \end{subfigure}%
        ~ %add desired spacing between images, e. g. ~, \quad, \qquad, \hfill etc.
        \begin{subfigure}[b]{0.5\textwidth}
                \includegraphics[width=\textwidth]{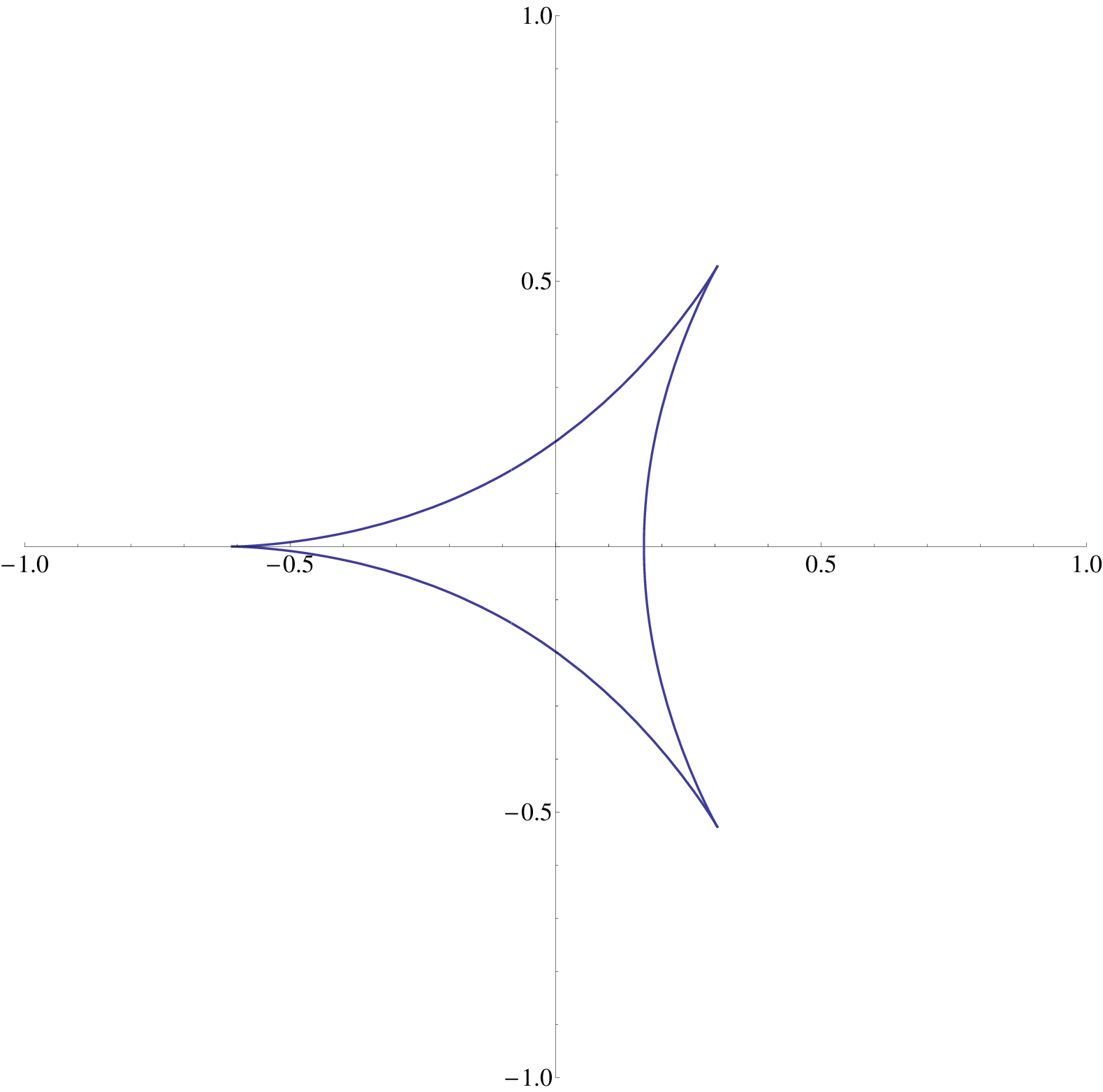}
                \caption{}
                \label{fig:k=2.0}
        \end{subfigure}

        \caption{Closed string profiles with $n=3$ for various values of $\varkappa$.
        The parameter values are a)~$(\varkappa,\omega) = (0.0001,1.05)$,
        b)$(\varkappa,\omega) = (10.0,22.0)$.}\label{fig:spikes2}
\end{figure}
\begin{figure}[!]
        \centering
        \begin{subfigure}[b]{0.5\textwidth}
                \includegraphics[width=\textwidth]{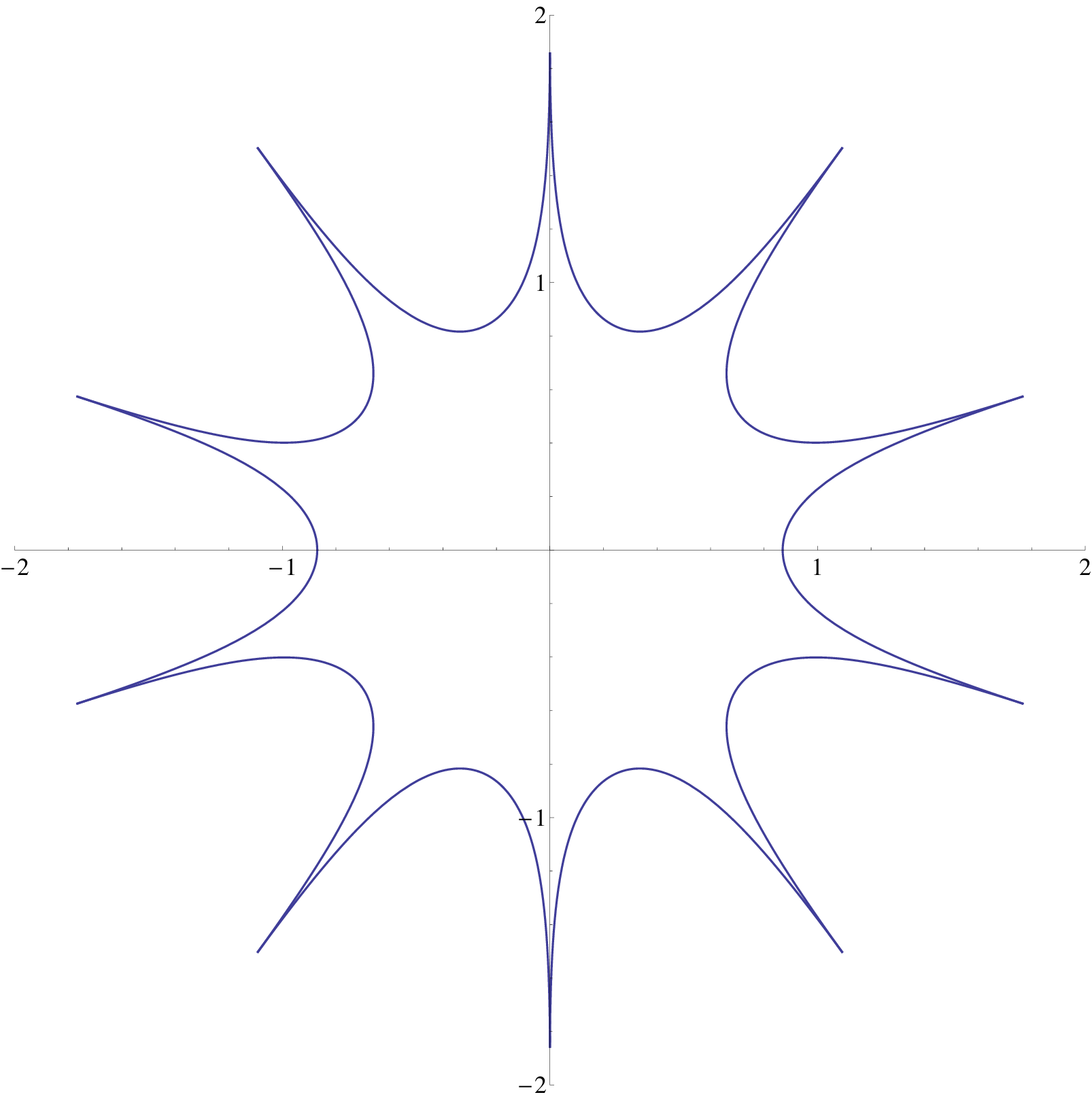}
                \caption{}
                \label{fig:k=10knotted}
        \end{subfigure}%
        ~ %add desired spacing between images, e. g. ~, \quad, \qquad, \hfill etc.
        \begin{subfigure}[b]{0.5\textwidth}
                \includegraphics[width=\textwidth]{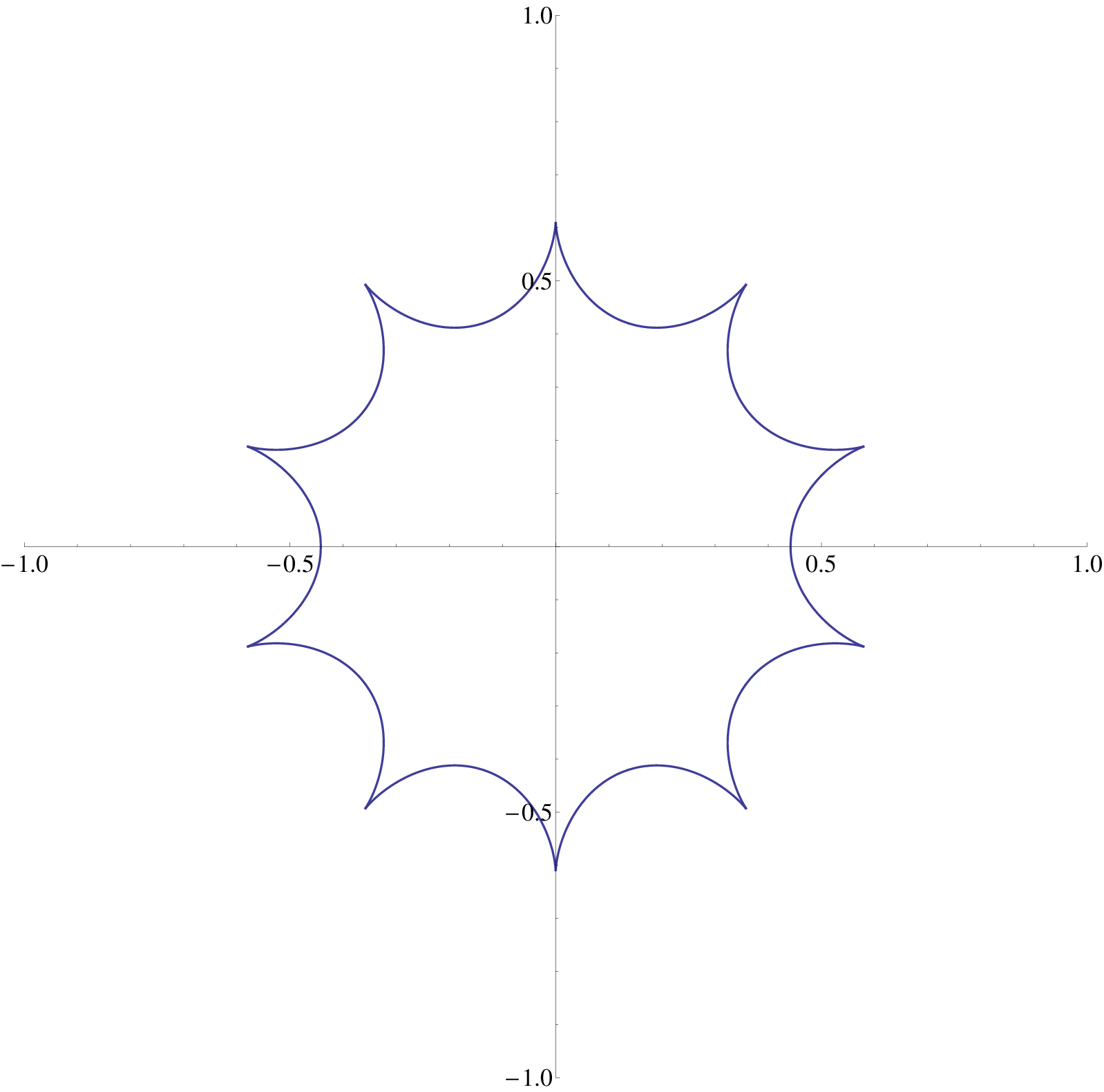}
                \caption{}
                \label{fig:k=10knotted3}
        \end{subfigure}

        \caption{Closed string profiles with $n=10$ for various values of $\varkappa$.
        The parameter values are a)~$(\varkappa,\omega) = (0.0001,1.05)$,
        b)$(\varkappa,\omega) = (10.0,22.0)$.}\label{fig:spikes3}
\end{figure}
We can fix the values $\chi_1^{\pm}$ by fixing the values of
$(\omega, \varkappa)$ by hand. As described in the previous
section, we can then extract the value of $\chi_0$ imposing the
closed string condition $\Delta\phi = \frac{2\pi}{2n}, \ \
n\in\mathbb{Z}$ on the string. This fixes all the parameters in
the calculation and the polar plot seemingly gives us the spiky
string solutions. Spiky strings for $n = 3$ and $n = 10$ are
plotted in figures (\ref{fig:spikes2}) and (\ref{fig:spikes3}) for
varying values of the parameters. Notice while the total profiles
look more or less the same, the nature for the single segment has
changed here. This can be understood by drawing tangents to the
single segments and comparing with the undeformed case. One can
actually verify that for the $t = \tau =constant$ the induced
mertrics on the worldsheet section for $AdS_3$ and
$(AdS_3)_\varkappa$ can be related by a conformal factor of $(1 +
\varkappa^2 \cosh^2\chi)^{-1}$. In that sense the spiky string
solutions on the worldsheet can also be related to each other. It
can also be noticed that as $\varkappa$ becomes smaller, the
spikes reach out more to the apparent boundary. However, we found
that for very high values of $\varkappa$ the closedness condition
does not admit a viable solution for $\chi_0$.

\begin{figure}
        \centering

                \includegraphics[width=0.8\linewidth]{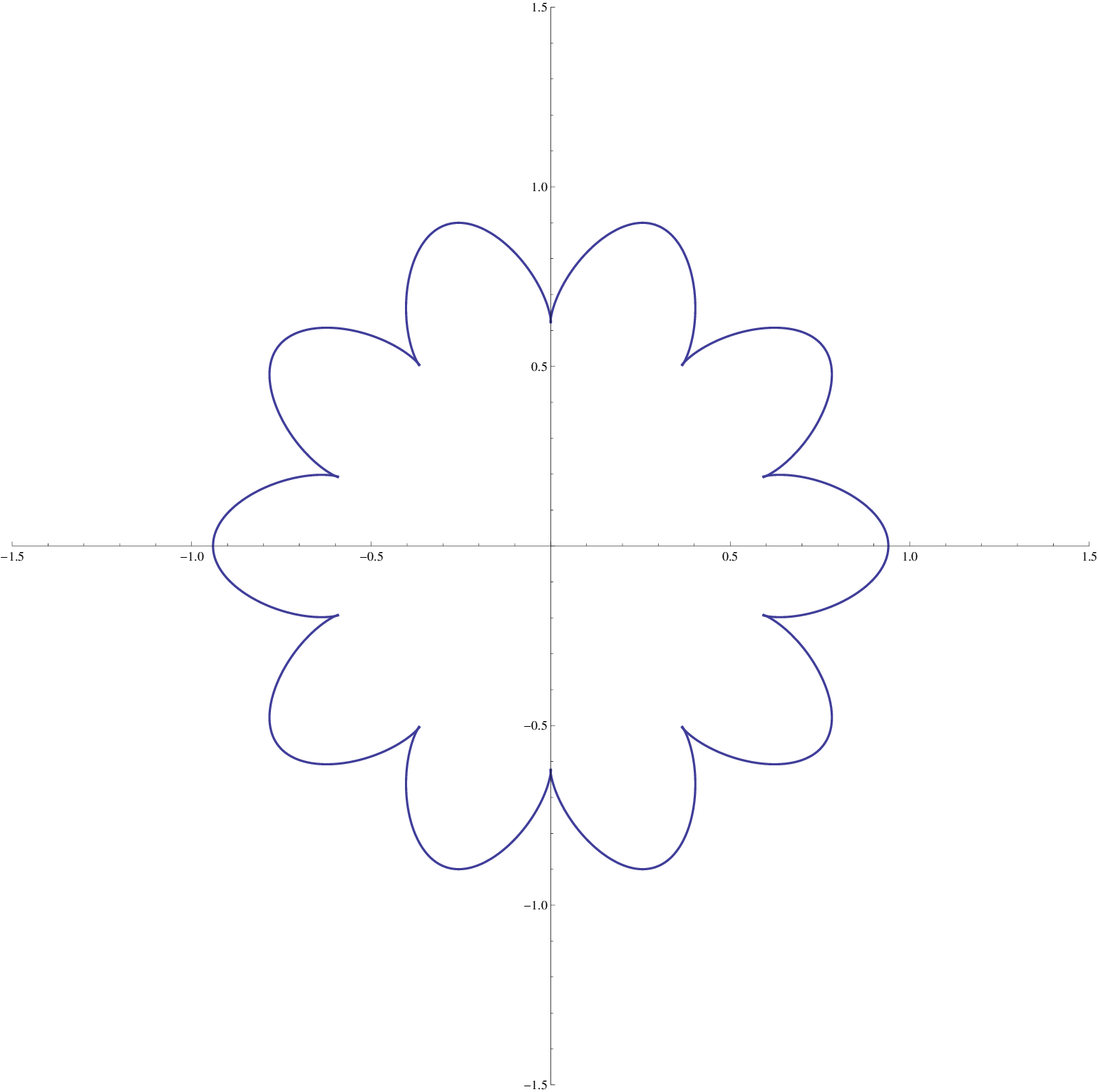}

        \caption{Spiky nature of closed strings forming towards the $(AdS)_\varkappa$ origin.
        Here the parameter values are $(\varkappa,\omega) = (2.0,4.7)$. }\label{fig:spikes5}
\end{figure}

A second kind of string solution is generated when we have
$\chi_0^{+} > \chi_1^{-}$, i.e. the so-called valleys form at
higher points than the spikes. For these values in the usual
undeformed problem, we should have got `dual' spikes
\cite{Mosaffa:2007ty} where spikes form facing the $AdS$ origin
instead of approaching the boundary. It is worth noting that the Energy-Spin
dispersion relation for such solutions is different than that of usual long spiky
strings. For such parameter values in our
deformed case we show the profiles remain unchanged. As an example
one can see figure (\ref{fig:spikes5}).
\subsubsection{A long string limit}
 For completeness, we look at the isometries of the background and construct the usual conserved
 quantities: energy $E$ and spin $S$.  The expression for the conserved charges 
 here lead to the following relation:
%\begin{equation}
%S= \frac{\omega}{2} \int_{\chi_0}^{\chi_1} \frac{\sinh{\chi}}{\cosh{\chi}}\frac{\sqrt{\sinh^2{2\chi}-(1+\varkappa^2\cosh^2{\chi})\sinh^2{2\chi_0}}}{(1+\varkappa^2\cosh^2{\chi})
%\sqrt{(1+\varkappa^2\cosh^2{\chi})\cosh^2{\chi}-\omega^2\sinh^2{\chi}}} d\chi
%\end{equation}
\begin{eqnarray}
\frac{E-\omega S}{n}
 &=& \frac{\sqrt{\lambda}\sqrt{1 + \varkappa^2}}{2\pi} \Biggr[\frac{\cosh{2r}-\cosh{2p}}{4(\cosh{2m}+\cosh{2r})}
\sqrt{\frac{\cosh{2q}-\cosh{2r}}{\cosh{2p}-\cosh{2s}}}
\mathbb{F}(\frac{\pi}{2},
\nu) \  \nonumber \\
&+& \frac{\cosh{2r}-\cosh{2s}}{4\sqrt{(\cosh{2p}-\cosh{2s})(\cosh{2q}-\cosh{2r})}}
\Pi(\zeta,
\frac{\pi}{2},\nu)  \  \nonumber \\
&+&\frac{
\frac{(\cosh{2p}+\cosh{2m})(\cosh{2q}+\cosh{2m})}{(\cosh{2r}+\cosh{2m})
(\cosh{2s}+\cosh{2m})}}{\sqrt{(\cosh{2p}-\cosh{2s})(\cosh{2q}-\cosh{2r})}}
\Pi(\Gamma,\frac{\pi}{2},\nu)\Biggr] \ ,
\end{eqnarray}
where we have,
\begin{eqnarray}
m &=& \frac{1}{2}\cosh^{-1}{(1+\frac{2}{\varkappa^2})}\  \ , \nonumber\\
\Gamma &=& \frac{(\cosh{2m}+\cosh{2r})(\cosh{2q}-\cosh{2s})}{(\cosh{2q}-\cosh{2r})
(\cosh{2m}+\cosh{2s})} \ \ , \nonumber\\
\zeta &=& \frac{\cosh{2q}-\cosh{2s}}{\cosh{2q}-\cosh{2r}}
\end{eqnarray}
The other symbols were defined before and $n$ is the number of
spikes as usual. Now, the above one is a complicated expression in
terms of various elliptic functions, which prove to be quite tough
to expand in particular limits. Also, we can see from
(\ref{chiplus}) that to get a `long' string limit of such
solutions ($\chi_1^{-} \rightarrow \infty$) with a fixed $\omega$,
one must apparently take $\varkappa\rightarrow 0$. We refrain from
discussing this from this point of view and present an alternative
analysis up to first order. 

To take the $\chi_1^{-} \rightarrow \infty$, we start with a redefinition of the coordinates
with $\sinh^2\chi_1^{-} = z$ and
\begin{equation}
 \cosh 2\chi_1^{-} = 2z + 1,~~~~~~ \sin u = \frac{\sinh \chi}{\sinh \chi_1^{-}}.
\end{equation}
Now with the redefinitions we can solve for $\omega$ from \ref{chiplus} to get
\begin{equation}
 \omega = \frac{\sqrt{1+z}\sqrt{1+(z+1)\varkappa^2}}{\sqrt{z}} \ ,
\end{equation}
where we ignore the -ve solution. Now since we have $\chi_1^{+}
>\chi_1^{-}$, we can take the following assumption
\begin{equation}
 \frac{\sinh^2\chi_1^{-}}{\sinh^2\chi_1^{+}} = 1- \delta ~~~~~~ \delta<<1.
\end{equation}
Now it can be shown that the $\delta<<1$ is necessary to take the long string limit.
Solving the above equation with the definitions of $\chi_1^{\pm}$ we can find that
\begin{equation}
 \sinh^2\chi_1^{-} = z = \frac{\sqrt{1 + \varkappa^2}}{\varkappa} + \mathcal{O}(\delta),
\end{equation}
which in turn gives
\begin{equation}
 \omega = \varkappa + \sqrt{1 + \varkappa^2} + \mathcal{O}(\delta^2).
\end{equation}
In what follows we will work in the leading order in $\delta$. Now, since the charges are
integrated from $\chi_0^{+}$ to $\chi_1^{-}$, we will need the condition $\chi_1^{-}>>\chi_0^{+}$
for a long string stretching upto the singularity surface which means,
\begin{equation}
 \sin u\mid_{\chi_0^{+}} \approx 0.
 \end{equation}
Now this can be achieved by taking $\chi_0\rightarrow 0$, which is
also evident from the Figure (\ref{fig:Rootlower}). Under these
considerations we can write the Energy $E$ as follows,
\begin{eqnarray}
 \frac{E}{n} &\simeq& \frac{\sqrt{\lambda}\sqrt{1 + \varkappa^2}}{\pi}
  \int_{\chi_0^{+}}^{\chi_1^{-}} \frac{\cosh^2{\chi}}
{\sqrt{(1+\varkappa^2\cosh^2{\chi})\cosh^2{\chi}-\omega^2\sinh^2{\chi}}} d\chi \ \nonumber\\
&=&
\frac{\sqrt{\lambda}}{\pi}\sqrt{z}\int_{0}^{\frac{\pi}{2}}\frac{\sqrt{1+z
\sin^2 u}} {\sqrt{1-(1-\delta)\sin^2 u}} du \ ,
\end{eqnarray}
where we have used the fact that
\begin{equation}
 \sinh^2 \chi_1^{+}~\sinh^2 \chi_1^{-} = \frac{1+\varkappa^2}{\varkappa^2}.
\end{equation}
Similarly we can write the spin of the spike as
\begin{equation}
 \frac{S}{n} =  \frac{\sqrt{\lambda}}{\pi}\sqrt{z}\omega\int_{0}^{\frac{\pi}{2}}
 \frac{z \sin^2 u}{(\sqrt{1-(1-\delta)\sin^2 u})(\sqrt{1+ z \sin^2 u})(1 + \varkappa^2(1+ z \sin^2 u))} du.
\end{equation}
While finding out the charges we have used that for a spiky strings with $n$ cusps,
\begin{equation}
 \int d\sigma = 2n \int_{\chi_0^{+}}^{\chi_1^{-}}\frac{d\chi}{\chi\prime} = 2\pi.
\end{equation}
Now due to the presence of the factor $\sqrt{1-(1-\delta)\sin^2 u}$ in denominator, the charges will diverge
in the upper limit.
We then choose the integration contour carefully and write the above two integrals in the form
\begin{equation}
 \int_{0}^{\frac{\pi}{2}} I =  \int_{0}^{\frac{\pi}{2} - 2\delta^{\frac{1}{4}}} I
 + \int_{\frac{\pi}{2} - 2\delta^{\frac{1}{4}}}^{\frac{\pi}{2}} I.
\end{equation}
For the first part of the  integration we can expand the integrals in orders of $\delta$ as follows
\begin{eqnarray}
 I_{1E} &=&  \frac{\sqrt{\lambda}}{\pi}\sqrt{z}\Biggr[\int_{0}^{\frac{\pi}{2}}\frac{\sqrt{1+z \sin^2 u}}{\cos u}
 + \mathcal{O}(\delta)\Biggr] du \ , \nonumber\\
I_{1S} &=&  \frac{\sqrt{\lambda}}{\pi}\sqrt{z}\omega
\Biggr[\int_{0}^{\frac{\pi}{2}}\frac{z \sin^2 u}{\cos u(\sqrt{1+ z
\sin^2 u})(1 + \varkappa^2(1+ z \sin^2 u))} +
\mathcal{O}(\delta)\Biggr] du \ . \nonumber \\
\end{eqnarray}
The charges now look analogous to the ones evaluated in \cite{Kameyama:2014vma}.
We can evaluate the  above integrals and expand them out in the form
\begin{eqnarray}
 \int_{0}^{\frac{\pi}{2} - 2\delta^{\frac{1}{4}}} I_{1E} = \frac{\sqrt{\lambda}}{\pi}\sqrt{z}
 \Biggr[&-&\sqrt{z}\sinh^{-1}\sqrt{z} + \frac{1}{4}\sqrt{1+z}~\ln \biggr( \frac{(1+z)^2}{\delta} \biggr)
 \ \nonumber\\
 &+&\frac{(2z-1)}{3\sqrt{1+z}}\sqrt{\delta} + \mathcal{O}(\delta) \Biggr]
 \ \nonumber
\end{eqnarray}
\begin{eqnarray}
 \int_{0}^{\frac{\pi}{2} - 2\delta^{\frac{1}{4}}} I_{1S} &=& \frac{\sqrt{\lambda}}{\pi}\sqrt{z}\omega
 \Biggr[- \frac{\sqrt{z}\sqrt{1+ \varkappa^2}}{1+(1+z)\varkappa^2}\tanh^{-1}\biggr(\frac{\sqrt{z}}
 {\sqrt{1+z}\sqrt{1+\varkappa^2}}\biggr) \ \nonumber\\
 &+& \frac{1}{4} \frac{z~\ln \biggr( \frac{(1+z)^2}{\delta} \biggr)}{\sqrt{1+z}(1+(1+z)\varkappa^2)}
 -\frac{z(4z^2\varkappa^2 - 5(1+\varkappa^2) - z(2+\varkappa^2))}{3(1+z)^{3/2}(1+(1+z)\varkappa^2)^2}\sqrt{\delta}
 + \mathcal{O}(\delta) \Biggr] \ \nonumber
\end{eqnarray}
Now the second part of the integral can be evaluated using a midpoint rule to be written as
\begin{eqnarray}
 I_{2E} &=&  \frac{\sqrt{\lambda}}{\pi}\sqrt{z}\Biggr[2\sqrt{1+z}-
 \frac{(2z-1)}{3\sqrt{1+z}}\sqrt{\delta} + \mathcal{O}(\delta) \Biggr] \ \nonumber\\
 I_{2S} &=& \frac{\sqrt{\lambda}}{\pi}\sqrt{z}\omega \Biggr[\frac{2z}{\sqrt{1+z}(1+(1+z)\varkappa^2)}
  + \frac{z(4z^2\varkappa^2 - 5(1+\varkappa^2) - z(2+\varkappa^2))}{3(1+z)^{3/2}(1+(1+z)\varkappa^2)^2}\sqrt{\delta}
 + \mathcal{O}(\delta) \Biggr] \ \nonumber
\end{eqnarray}
Now one can see that when we add everything up, $E$ and $S$ does not contain any $\mathcal{O}(\sqrt{\delta})$
term. Then we can combine the above expressions and use the expressions for $z$ and $\omega$ to find
\begin{eqnarray}
 \frac{E - (\varkappa + \sqrt{1 + \varkappa^2})S}{n} &=& \frac{\sqrt{\lambda}}{\pi}
 \frac{\sqrt{1 + \varkappa^2}}{\varkappa}\Biggr[-\sinh^{-1}\biggr(\sqrt{
 \frac{\sqrt{1 + \varkappa^2}}{\varkappa}}\biggr)
 \ \nonumber\\
 &+& (\varkappa + \sqrt{1 + \varkappa^2})\tanh^{-1}\biggr(\frac{1}{\sqrt{1+\varkappa^2}
 \sqrt{1+\frac{\varkappa}{\sqrt{1+\varkappa^2}}}}\biggr) + \mathcal{O}(\delta)
 \Biggr]  \ ,\nonumber \\
\end{eqnarray}
Now as in \cite{frolov} we define the changed variables
\begin{equation}
 w_0 \equiv \sqrt{1+\frac{\varkappa}{\sqrt{1+\varkappa^2}}},
 ~~~~k_0 \equiv \sqrt{1-\frac{\varkappa}{\sqrt{1+\varkappa^2}}} \
 .
\end{equation}
With a little manipulation, we can rewrite the key expressions as
\begin{eqnarray}
-\sinh^{-1}\biggr(\sqrt{
 \frac{\sqrt{1 + \varkappa^2}}{\varkappa}}\biggr) &=& \frac{1}{2}
 \ln \Biggr[\frac{w_0 - 1}{w_0 + 1} \Biggr] \ ,
 \ \nonumber\\
 \tanh^{-1}\biggr(\frac{1}{\sqrt{1+\varkappa^2}
 \sqrt{1+\frac{\varkappa}{\sqrt{1+\varkappa^2}}}}\biggr)
 &=& \frac{1}{2} \ln \Biggr[\frac{1 + k_0}{1 - k_0} \Biggr] \ .
 \end{eqnarray}
Then we are led to the dispersion relation of the exact form as in \cite{frolov, Kameyama:2014vma}, 
provided we take $n=2$, i.e. the solution reduces to that of a GKP folded
string solution. The final expression is of the form,
\begin{equation}
 \frac{E - \frac{w_0}{k_0}S}{n} = \frac{\sqrt{\lambda}}{2\pi}
 \frac{\sqrt{1 + \varkappa^2}}{\varkappa}\Biggr[\frac{w_0}{k_0}\ln \biggr[\frac{1 + k_0}{1 - k_0} \biggr]
 + \ln \biggr[\frac{w_0 - 1}{w_0 + 1} \biggr]  + \mathcal{O}(\delta) \Biggr].
\end{equation}
From the above expression one can note that in the $\varkappa \rightarrow 0$ limit the
right hand side actually diverges. To take the other extreme limit $\varkappa \rightarrow \infty$
we define the scaled quantities
\begin{equation}
 \mathcal{E} = \frac{E}{\frac{\sqrt{\lambda}}{2\pi}\sqrt{1+\varkappa^2}}~~~~
 \mathcal{S} = \frac{S}{\frac{\sqrt{\lambda}}{2\pi}\sqrt{1+\varkappa^2}}.
\end{equation}
Now in the $\varkappa \rightarrow \infty$ limit we perform the rescalings
$\mathcal{E} \rightarrow \mathcal{E}$ and $\mathcal{S} \rightarrow \mathcal{S}/\varkappa$
to keep the charges finite. The dispersion relation here looks like,
\begin{equation}
 \frac{\mathcal{E} - 2\mathcal{S}}{n} = 2\sqrt{2} + \ln(3 - 2\sqrt{2}) + \mathcal{O}(\delta).
\end{equation}
Remember, in this limit the string sigma model is related to that of $dS_3\times H^3$ by T-duality.
By making the maximal deformation, we can actually interpolate between positive and negative curvature
target spaces. But the ramification of these string solutions are yet to be explored fully.

\section{Summary and conclusion}
In this work we have presented `spiky' string solutions in
$\varkappa$ deformed $AdS$ in details. After a brief discussion on
the original solutions in undeformed $AdS$ we find the
corresponding string profiles in the one parameter deformed model
and visualise them graphically. We have used here the new type of
coordinate system where only the spacetime inside the singularity
surface is described. Both polyakov strings with the most general
type of of embedding ansatz and Nambu-Goto type simple strings
were discussed and classified here. It was found that in the first
case for the Polyakov strings the spikes seem to `open up' or
become parallel as $\varkappa$ is increased, the  spike solutions
of \cite{Kruczenski:2004wg} remain more or less similar as far as
the cusp structures are concerned. We have also found out the
`Dual' spike soltions forming towards the $(AdS)_\varkappa$
origin. All these new solutions conform to the fact that classical
string solutions in this background can not be stretched to the
actual boundary. In such cases the usual $AdS/CFT$ dictionary
might not work completely, but inside the singularity surface we
hope to make some sense out of the usual string solutions.
Although the nature of these solutions have to be understood in a
more rigorous way to justify this claim.

There are many problems still to be addressed here. One can look
at the holographic entanglement entropy in this background to find
out whether the usual area law holds or not. The nature of the
singularity surface may be investigated in even more details
using the full supergravity solution. The most fascinating
question that remains unanswered here is the dual gauge theory
side which might be able to shed light on the behaviour of these
classical strings. The dual theory, if any, is expected to be
highly nonlocal and requires careful investigation.
\section*{Acknowledgement}
The authors are indebted to Prof. S. Pratik Khastgir for elaborate discussions and comments
on the work. AB would also like to thank the organisers of 9th Asian Winter School on Strings, Particles,
and Cosmology for kind hospitality at Busan, South Korea, where initial stages of this work were done.

\end{document}